\documentclass{article}

\usepackage{arxiv}
\usepackage[utf8]{inputenc} % allow utf-8 input
\usepackage[T1]{fontenc} % use 8-bit T1 fonts
\usepackage{hyperref}       % hyperlinks
\usepackage{url}            
\usepackage{booktabs}       % professional-quality tables
\usepackage{amsfonts}       % blackboard math symbols
\usepackage{nicefrac}       % compact symbols for 1/2, etc.
\usepackage{microtype}      % microtypography
\usepackage{lipsum}
\usepackage{graphicx}
\usepackage{longtable}
\usepackage{float}
\usepackage{caption}
\usepackage{subcaption}
\graphicspath{ {./images/} }
\usepackage[style=apa]{biblatex} 
\title{Understanding Heterogeneity in Adaptation to Intermittent Water Supply: Clustering  Household Types in Amman, Jordan}

\author{
 Shreyas Gadge \\
  Institute for Biodiversity and Ecosystem Dynamics\\
  University of Amsterdam\\
  Amsterdam, NL \\
  \texttt{s.r.gadge@uva.nl} \\
   \And
 Vítor V. Vasconcelos \\
  Computational Science Lab, Informatics Institute\\
  Polder center, Institute for Advanced Study\\
  University of Amsterdam\\
  Amsterdam, NL \\
  \texttt{v.v.vasconcelos@uva.nl} \\
  \And
 André de Roos \\
  Institute for Biodiversity and Ecosystem Dynamics\\
  University of Amsterdam\\
  Amsterdam, NL \\
  Santa Fe Institue\\
  Santa Fe, New Mexico, USA\\
  \texttt{A.M.deRoos@uva.nl} \\
  \And
 Elisabeth H. Krueger \\
  Institute for Biodiversity and Ecosystem Dynamics\\
  University of Amsterdam\\
  Amsterdam, NL \\
  \texttt{e.h.krueger@uva.nl} \\ 
}

\begin{document}
\maketitle
\begin{abstract}
More than a billion people around the world experience intermittence in their water supply, posing challenges for urban households in Global South cities. An intermittent water supply (IWS) system prompts water users to adapt to service deficits which entails coping costs. Adaptation and its impacts can vary between households within the same city, leading to intra-urban inequality. Studies on household adaptation to IWS through survey data are limited to exploring income-based heterogeneity that do not account for the multidimensional and non-linear nature of the data. There is a need for a standardized methodology for understanding household responses to IWS that acknowledges the heterogeneity of households characterized by sets of multiple underlying factors and that is applicable across different settings. Here, we develop an analysis pipeline that applies hierarchical clustering analysis (HCA) in combination with the Welch-two-sample t-test on household survey data from Amman, Jordan. We identify three clusters of households distinguished by a set of characteristics including income, water social network, supply duration, relocation and water quality problems and identify their group-specific adaptive strategies such as contacting the utility or accessing an alternate water source. This study uncovers the unequal nature of IWS adaptation in Amman, giving insights into the link between household characteristics and adaptive behaviors, while proposing a standardized method to reveal relevant heterogeneity in households adapting to IWS. 
\end{abstract}

% keywords can be removed

\section{Introduction}\label{sec1}
Adaptation at the level of households and citizens is an essential component of social-ecological systems, as it complements government-led actions, dynamically responding to climate- and management-induced adversities while balancing socio-economic priorities \parencite{taberna_uncertainty_2023}. A pragmatic understanding of the adaptation process at household level can help to uncover vulnerability within specific social-ecological contexts and integrate it into better resource management \parencite{chapagain_studies_2025}.The heterogeneity of human adaptation has been studied in the context of water systems by exploring socio-economic and behavioral determinants of household vulnerability to droughts \parencite{oluwatayo_socioeconomic_2022, bryan_coping_2019}, intra-urban inequalities in drought responses \parencite{savelli_urban_2023}, and household adaptation to floods \parencite{taberna_uncertainty_2023, di_baldassarre_socio-hydrology_2013}. Here, we focus on uncovering heterogeneity in household adaptation to water scarcity, specifically in intermittent water supply systems.

Intermittent Water Supply (IWS) denotes piped water services with supply durations of less than 24 hours per day or less than 7 days per week  \parencite{sashikumar_modelling_2003}. More than a billion people around the world experience IWS \parencite{bivins_estimating_2017}, posing challenges for urban households in Global South cities \parencite{beard_water_2021}. While temporal supply disruptions may result from droughts or maintenance, more common in water-abundant and technologically advanced countries, intermittence is used as a measure to control demand and water use in cities with permanent water scarcity. This, also referred to as full-time intermittence \parencite{mcintosh_asian_2003}, is the focus here. Urban managers opt for IWS when services cannot keep pace with rapid urbanization, demand growth, and hydrological regime changes, among others \parencite{simukonda_intermittent_2018}. 

%However, IWS also strains water infrastructure, leading to high maintenance costs faced by the utility companies, which can reinforce intermittence. Conditions of IWS can form an undesirable reinforcing feedback loop, creating a "spiral of decline" \parencite{galaitsi_intermittent_2016}. For example, water access inequality caused by perceived unreliability of IWS leads to water users investing in private water infrastructure, such as storage facilities \parencite{vasquez_reliability_2012}. There is also evidence that investment in private water infrastructure generates inequality among users \parencite{de_marchis_analysis_2011}, which can further lead to private investment, making it a reinforcing loop within IWS. Water suppliers emphasize natural resource constraints underlying IWS, which may lead to an undervaluing of human drivers \parencite{galaitsi_intermittent_2016}, and give a skewed explanation to this complex problem. Therefore, recent studies suggest the importance of considering water users' practices and perceptions to address the causes and consequences of IWS and resulting inequalities \parencite{savelli_dont_2021} \parencite{obringer_improving_2022}. 

IWS prompts water users to adapt by accessing alternate sources, such as buying water delivered by tanker trucks or bottled water, or seeking help from others to accommodate the intermittence and unreliability of water supply. IWS and the adopted coping strategies can have negative impacts on the welfare of households, water quality, health, and time invested in water collection \parencite{huberts_making_2023, li_intermittentwater_2020, simukonda_intermittent_2018, majuru_how_2016}. To inform actions for improving the well-being of a growing urban population living with IWS, it is important to understand household adaptation to IWS and its underlying characteristics \parencite{simukonda_intermittent_2018}. 

Water insecurity among households can especially be concerning for particular groups within society \parencite{hoekstra_urban_2018}, such as racial minorities and those living in poverty conditions \parencite{chapagain_studies_2025}. Adaptation and its impacts can vary between households within the same city, owing to significant intra-urban inequality \parencite{krueger_resilience_2019, savelli_urban_2023, kromker_who_2008}. IWS amplifies inequitable access to water due to heterogeneous supply durations throughout the city \parencite{galaitsi_intermittent_2016,simukonda_intermittent_2018}. Previous studies on household responses to unreliable water supplies, which typically rely on survey data \parencite{beard_water_2021}, have focused on establishing a link between household income and time investment, alternative-source costs, and coping strategies such as storage capacities \parencite{pattanayak_coping_2005,katuwal_coping_2011,satpathy_intermittent_2022,cook_costs_2016,potter_contemporary_2010}. These studies show that the coping costs in response to IWS are unevenly distributed between income groups due to the difference in the types of strategies each group adopts and their associated costs \parencite{majuru_how_2016, chelleri_integrating_2015}. However, given the multidimensional nature of adaptation, resulting from socioeconomic and behavioral factors interacting with biophysical constraints, income or supply duration alone may not sufficiently explain the heterogeneity of household responses to IWS \parencite{taberna_uncertainty_2023, chapagain_studies_2025}.

Explorations of socioeconomic determinants of urban water supply insecurity have employed methods such as qualitative analyses of case studies \parencite{grasham_socio-political_2021,satur_social_2020,huberts_making_2023}, site-specific descriptive statistics \parencite{li_intermittentwater_2020,khan_beyond_2022}, or regression models \parencite{shah_variations_2023,li_intermittentwater_2020}. While these case studies highlight the importance of household-level factors in shaping water use and acknowledge the presence of social gradients and inequalities \parencite{grasham_socio-political_2021,satur_social_2020,al-kassab-cordova_spatial_2023}, their exploration of heterogeneity is often limited to reporting differences in household adaptation across individual variables, typically spatial location, income groups or supply duration \parencite{li_intermittentwater_2020,khan_beyond_2022,grasham_socio-political_2021}. Such studies of heterogeneity are limited to linear inequality measures such as the Gini coefficient or concentration curves \parencite{al-kassab-cordova_spatial_2023}, which effectively capture aggregate inequality of the distribution of a single variable but do not account for the interplay of several factors leading to group-specific insights. Thus, there is a need for a standardized methodology for understanding household responses to IWS that acknowledges the heterogeneity of households through underlying sets of characteristics and can be applied across different settings to enable meaningful comparisons of the distribution of adaptive strategies within a city \parencite{majuru_how_2016}.
\\ \\ 
Here, clustering methods can be a useful tool, which have been used to characterize household types according to water consumption patterns and use behaviors, as well as farmer households adapting to climate change \parencite{bryan_coping_2019,pilo_impacts_2021,maleksaeidi_discovering_2016,sattar_typology_2024,ioannou_exploring_2021,almulhim_segmentation_2024}. We apply clustering analysis to characterize household types and their potentially distinct adaption to IWS, using household survey data from a study conducted in Amman, Jordan's capital city, which has limited water availability and supplies water to its households intermittently \parencite{krueger_reframing_2025}. We use hierarchical clustering analysis (HCA; Section \ref{hca}) to characterize the heterogeneous household groups based on a combination of household-level characteristics. Then, we analyze the distribution of adaptive behaviours among these groups. In a study characterizing social-ecological system archetypes in Andalusia, Spain, HCA was used in combination with a random forest approach to characterize the importance of each indicator within a cluster \cite{pacheco-romero_data-driven_2022}. Here, we use HCA in combination with the Welch two-sample t-test (Section \ref{ttest}), which allows identifying significant features that distinguish a cluster from the rest of the dataset. It provides a deterministic, per-feature statistical interpretation that supports mechanistic understanding of the cluster. HCA is suitable for analyzing the distribution of adaptative behaviours to IWS, as it considers multiple underlying factors from high-dimensional data. The Welch two-sample t-test was used in a previous study that compared household clusters based on socioeconomic information from survey data in Kampala, Uganda \parencite{hemerijckx_upscaling_2020}. However, the combination of HCA and the Welch two-sample t-test has not been previously applied to study household adaptation within IWS. Our analysis is guided by the following research questions: 
\begin{itemize}
    \item What are the different types of household groups adapting to IWS, and which set of characteristics distinguishes them?
    \item Can we associate and identify specific adaptive behaviors with specific types of households?
\end{itemize}

The paper is organized as follows: Section \ref{sec2} highlights the challenges of water scarcity in Amman, section \ref{method} describes the methods used for analyzing the household survey data, section \ref{result} presents the emergent household clusters, their comparison and within-cluster heterogeneity, while section \ref{dc} discusses the results and limitations, and section \ref{cc} draws conclusions from our study.
\\ \\

\section{Case study}\label{sec2}
The Hashemite Kingdom of Jordan is an arid, mostly landlocked country with extremely limited natural water availability. Much of its water sources are  shared with neighboring countries. Its growing population, declining rainfall and increasing temperatures due to climate change exacerbate the country's water scarcity, challenging reliable drinking water supply \parencite{yoon_coupled_2021}. Jordan's political stability has attracted migrants from Syria, Palestine, Lebanon, and Iraq \parencite{potter_contemporary_2010}. While there are about 700,000 refugees officially registered in Jordan (UNCHR 2025), conflict-driven immigrant numbers by far exceed this number, as reflected by the history of conflicts repeatedly driving steep population increases over many decades  \parencite{potter_ever-growing_2009}.
%Jordan had over 1.3 million refugees for a population of 11 million in 2024 (acaps). 
Amman, the capital of Jordan, had an estimated population of over 4.9 million at the end of 2024 (Population Statistics Directorate, Jordan). In 2015, around the time when the data analyzed in this study was collected, the average per capita water supply was as low as 68 liters per day at household level \parencite{krueger_reframing_2025}. Amman has a sophisticated and expensive water supply infrastructure transferring water over distances of 350 km and pumping water over altitudes of 1000 m \parencite{krueger_reframing_2025}. 98\% of households in Amman are connected to the public water supply network; however, since 1987, the water supply has been rationed \parencite{potter_issues_2010}. As a result, Amman residents have to resort to several measures to increase the amount, duration, and safety of water supply, including the storage of water in rooftop tanks, buying water from tanker trucks or stores, or filtering water before use \parencite{potter_contemporary_2010}. The costs associated with these measures, both financial and time invested, amplify existing social inequalities. Amman's districts are characterized by a social divide with the eastern neighborhoods housing the urban poor, while the western and northern parts are inhabited by more affluent groups \parencite{potter_contemporary_2010}. Income variation in Amman has been linked to household water use and adaptive strategies under 'water stress' \parencite{potter_issues_2010} to differences in levels of water consumption, household expenditure on water, and storage capacities \parencite{potter_contemporary_2010}.

\section{Methods}\label{method}

\subsection{Data}\label{data}

We use secondary data collected in 2016 through a household survey in Amman as part of a study on characterizing social-ecological-technological system interactions and uncertainties \parencite{krueger_reframing_2025}. Exploring the resilience of Amman's water supply, 300 households were surveyed, based in 10 neighbourhoods (water supply districts) with different socio-economic characteristics, supply durations, and infrastructure reliability (pipe bursts), and covering a cross-section of the city from North to South and from East to West \parencite{krueger_reframing_2025}. The questions centered around the following themes: 1) water supply reliability, 2) water quality, 3) adaptation to water service deficits, 4) challenges to the long-term, reliable urban water supply and potential solutions to these challenges, 5) the respondents' contacts and interactions with others around water-related issues, and 6) demographics. This dataset includes information about several coping strategies and factors underlying household adaptation to IWS, which enabled us to evaluate the heterogeneity of household characteristics and adaptation behavior in response to water supply deficits. While the data was published in a recent paper focusing on aggregate household perceptions of the resilience of Amman's water system, the data has not been analyzed in a disaggregated way to understand heterogeneous household characteristics and responses to IWS, which is the focus here.

\subsection{Data Pre-Processing}\label{datap}
We chose variables from the survey data that represent household-level information about: (1) water supply reliability, duration, and coverage of household demand, (2) household size, income, and migration status, (3) adaptation to water service deficits, (4) problems experienced by households due to IWS. \\ \\ We removed one data point with a household size equal to 0 (a restaurant establishment), resulting in a total of 299 data points. The dataset includes two types of categorical variables: (1) variables pre-categorized (or binary) in the questionnaire (for example, delay length, storage type, and alternate source frequency), which remain unchanged; (2) response variables categorized post-survey making them ordinal variables (for example income and storage size).
\\ \\
Missing values and ambiguous responses (for example, 'I don't know') in the dataset were handled using domain knowledge and statistical imputation. The response was categorized by stating assumptions for a missing response or inferring the category using response to a corresponding question or replacing with the numerical mean across all responses. This imputation assumes missingness is random or not strongly biased by unmeasured factors. We computed and used income, storage size, and water bill as per capita values, ensuring comparability across households of different sizes. In the variable 'water social network,' which asked for names and relations of 5 people to contact in case of issues with water, we removed 'utility' from the responses, as it was already considered as a delay strategy ('contact utility'). For details, see appendix \ref{amm}.
\\ \\ 
We divided the data between variables that represent household characteristics ('characteristic variables') and variables that represent adaptive behaviors in response to water service deficits ('outcome variables'). We identified 18 characteristic variables (6 binary, 9 ordinal, and 3 nominal) and 4 outcome variables (1 binary, 1 ordinal, and 2 nominal), which can be found in column 1 of Tables 1 and 2, respectively. 
\\ \\ 
All variables were pre-processed by (1) one-hot encoding (OHE) of nominal variables and (2) Z-score standardization of all variables. OHE creates binary indicator variables for each category of the nominal variable, making it suitable to analyze alongside ordinal variables (as long as it does not overly fragment the data or induce severe multicollinearity). Z-score standardization ensures that variables with different scales, category ranges, and measurement units contribute equally to hierarchical clustering by transforming them to a common scale with zero mean and unit variance, preventing dominant influence from higher-magnitude features. Hierarchical Clustering Analysis (HCA) (\ref{hca}) was subsequently performed on the dataset and Welch two-sample t-test (\ref{ttest}) was used to identify cluster-wise significant variables We performed correlation analysis (\ref{acorr}) and principal component analysis (\ref{apca}) on the processed dataset; however, we did not observe significant correlations between chosen variables and the first two principal components (PC-1 and PC-2) only explained about 17.5\% and 15\% variance in the dataset respectively. 

\begin{figure}[H]
    \centering
    \includegraphics[width=10cm]{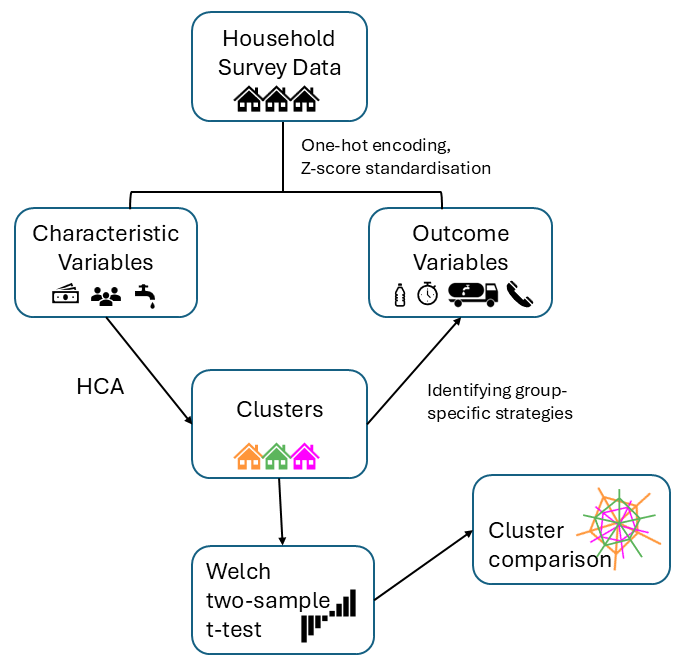}
    \caption{Overview of steps of analysis on household survey data}
    \label{fig:1}
\end{figure}

Figure \ref{fig:1} illustrates the overview of the steps of analysis. 

\subsection{Hierarchical Cluster Analysis}\label{hca}
Hierarchical Cluster Analysis (HCA) was performed on the characteristic variables only. The outcome variables were used later to understand cluster-specific behaviors. HCA is an agglomerative approach to clustering a dataset in the following steps:
For cluster $C_{i}$ with $n_{i}$ number of households, the Sum of Squared Errors (SSE) is calculated as,
\begin{equation}
    SSE_{C_{i}} = \sum_{j=1}^{n_{i}}\mid\mid O_{j} - \bar O\mid\mid^{2}
\end{equation}
where $O_{j}(j = 1,2...,n_{j})$ is the $j^{th}$ observation in the cluster, $\bar O$ is the mean of all observations in the cluster, and $\mid\mid O_{j} - \bar O\mid\mid^{2}$ is the squared Euclidean distance between $O_{j}$ and $\bar O$. HCA begins with $n$ clusters of $n$ data points and, in the first step, $\Delta SSE$ is calculated for all possible cluster combinations. For instance, if clusters $A$ and $B$ are to be merged into a new cluster $C$, $\Delta SSE$ (change in SSE) is calculated as, 
\begin{equation}
    \Delta SSE = SSE_{C} - (SSE_{A} + SSE_{B})
\end{equation}
According to Ward's criterion, the two clusters with the smallest $\Delta SSE$ are merged into a new cluster. A dendrogram records the merging history, with the x-axis representative of all samples and the y-axis representing the clustering history based on the linkage-distance ($\sqrt{2\Delta SSE}$). The number of clusters is determined by placing a line at a specific linkage distance in the dendrogram (the 'phenon line'). The location of the phenon line in the dendrogram is chosen to optimize cluster separation while preserving meaningful group structures. The number of branches that the phenon line crosses in the dendrogram is the number of clusters in the sample. We arrive at the final number of clusters in this study by testing different cut-off levels in the dendrogram to obtain a comprehensible picture of the diversity of households adapting to IWS, while interpreting the significant characteristics (using Welch two-sample t-test) based on knowledge of the study area, as is also followed in the study of social-ecological archetypes within Andalusia, Spain \cite{pacheco-romero_data-driven_2022}. In this study, we set the number of clusters to three to balance internal cluster cohesion and interpretability. Other methods to determine the number of clusters (e.g., silhouette scores, elbow methods as used in \parencite{liu_using_2021}) can be used, but in our case resulted in lower interpretability.

\subsection{Welch two-sample $t$-test}\label{ttest}
Once the number of clusters is decided, the significance of each variable (characteristic and outcome) within a cluster is determined through a Welch two-sample $t$-test. Unlike the standard $t$-test, the Welch test does not assume equal variances across groups, making it particularly robust for comparing the heterogeneous household clusters. This test compares the mean value of each variable in the cluster to the means of these variables in the rest of the dataset. The $t$-value is a standardized measure of the difference in means between a cluster and the rest of the dataset relative to the variation in the data. We calculate $t$-values for each feature in the dataset, for all clusters. A positive or negative $t$-value signifies whether the feature's mean is higher or lower, respectively, for the cluster compared to the rest of the dataset. The magnitude of the t-value indicates the strength and significance of the difference. Features with large absolute t-values are considered key contributors to explaining the distinct characteristics of a cluster. The $t$-test fails for a feature when there is no variance in either of the groups. This means that all households within one of the groups (cluster or rest of the dataset) have the same response for the feature. Since the $t$-test relies on differences in means relative to variance, the lack of variability in one of the groups prevents the calculation of a meaningful $t$-statistic. Such features are separately recorded in Table 3. To account for multiple hypothesis testing (across multiple features within a cluster), we applied the Benjamini-Hochberg correction to control the false discovery rate (FDR), using the multipletests function from the statsmodels package in python. This corrects the p-values associated with each feature, and if they are lower than $\alpha = 0.05$, the feature is significant in describing the cluster relative to rest of the dataset.

\newpage

\section{Results}\label{result}
\begin{figure}[H]
    \centering
    \includegraphics[width=15cm]{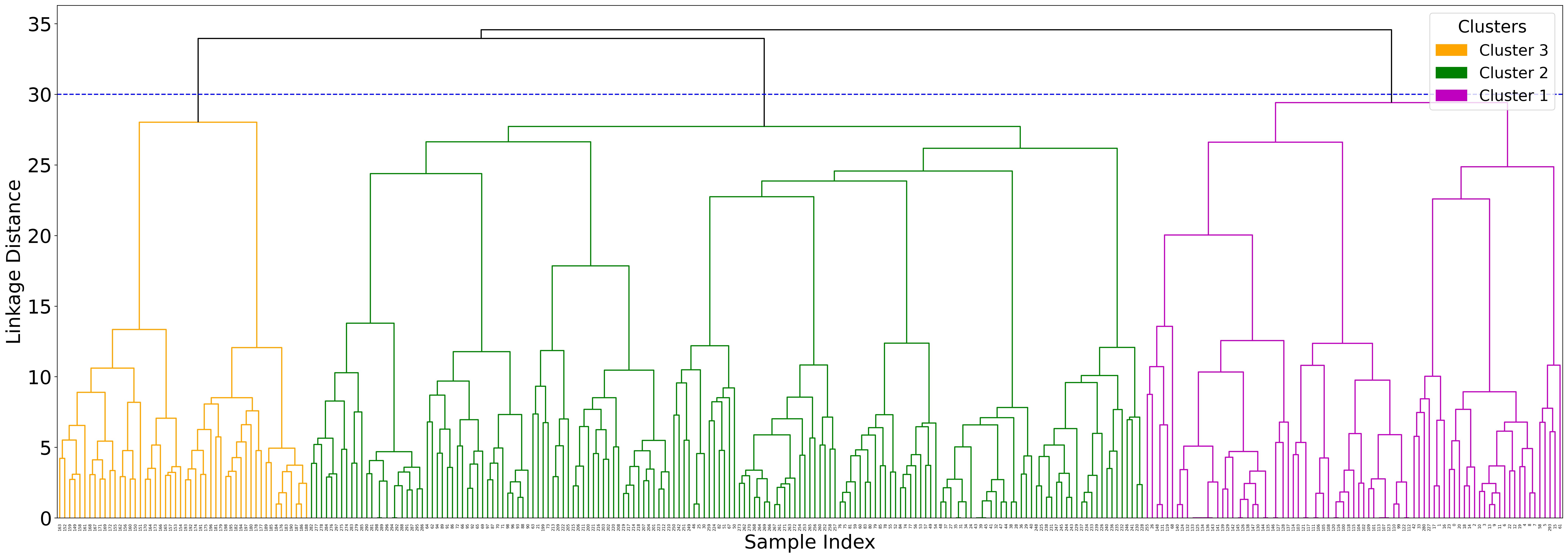}
    \caption{HCA Dendrogram with cluster merging history of household survey data from Amman. Cut at a linkage distance of 30, it results in three distinct clusters (from the right: magenta - cluster 1, green - cluster 2, orange - cluster 3).}
    \label{fig:2}
\end{figure}

The HCA performed on the characteristic variables resulted in a dendrogram that showcases the hierarchical structure of the data. The dendrogram shown in Figure \ref{fig:2} was cut at a linkage distance of 30, yielding three distinct clusters of households with variable size (cluster 1: N=83, cluster 2: N=166, cluster 3: N=50). We can also see the cluster merging history for the dataset from the dendrogram, where each individual household is its own cluster (at the bottom), iteratively merging until the whole dataset is one cluster (top), recording the linkage distance on the y-axis (Figure \ref{fig:2}). Table 1 records the count and proportion of households in each cluster and in the overall dataset for each characteristic variable, while Table 2 records the same for outcome variables indicating adaptive behaviors. The cumulative percentage for the overall data and each cluster may exceed 100 for outcome variables delay strategy (DS) and alternate source (AS), as multiple responses were possible there. 

\begin{longtable}
{p{0.13\textwidth}p{0.17\textwidth}p{0.13\textwidth}p{0.13\textwidth}p{0.13\textwidth}p{0.13\textwidth}p{0.13\textwidth}}
\toprule
\caption{Characteristic Variables: Cluster-wise proportion of households across categories \\\centering ($N = 299$, $n_{1} = 83$, $n_{2} = 166$, $n_{3} = 50$)}\label{tab:1}\\ \\
Characteristic Variable (unit)&  Category (response) & Overall Count (\%) & Cluster 1 Count (\%) & Cluster 2 Count (\%) & Cluster 3 Count (\%) \\
\midrule
\endfirsthead
\toprule
Characteristic Variable (unit)&  Category (response) & Overall Count (\%) & Cluster 1 Count (\%) & Cluster 2 Count (\%) & Cluster 3 Count (\%)\\
\midrule
\endhead
\midrule
\endfoot 
\bottomrule
\endlastfoot
Supply  &      1 ($\leq$ 1) &       49 (16.39\%) &           3 (3.61\%) &         28 (16.87\%) &          18 (36.0\%) \\
 Duration  &      2 (>1 - 3) &      146 (48.83\%) &         37 (44.58\%) &         95 (57.23\%) &          14 (28.0\%) \\
(days/week) &      3 (>3 - 6) &        21 (7.02\%) &         10 (12.05\%) &           9 (5.42\%) &            2 (4.0\%)\\
 &      4 (>6 - 7) &       83 (27.76\%) &         33 (39.76\%) &         34 (20.48\%) &          16 (32.0\%) \\
 \\Delay Length &      1 (no delay) &      132 (44.15\%) &          20 (24.1\%) &         77 (46.39\%) &          35 (70.0\%)\\
(days) &      2 ($\leq$1) &      112 (37.46\%) &         54 (65.06\%) &         49 (29.52\%) &           9 (18.0\%) \\
 &      3 (2) &       36 (12.04\%) &           8 (9.64\%) &         23 (13.86\%) &           5 (10.0\%)\\
 &      4 ($\geq$3) &        19 (6.35\%) &            1 (1.2\%) &         17 (10.24\%) &            1 (2.0\%)\\
\\Storage Size &      1($\leq$ 1) &      223 (74.58\%) &         54 (65.06\%) &        133 (80.12\%) &          36 (72.0\%) \\
($m^{3}$/capita) &      2 (>1 - 3) &       62 (20.74\%) &         25 (30.12\%) &         26 (15.66\%) &          11 (22.0\%) \\
 &      3 (>3) &        14 (4.68\%) &           4 (4.82\%) &           7 (4.22\%) &            3 (6.0\%)\\
 \\ Storage  &      1 (rooftop, only)&      234 (78.26\%) &         66 (79.52\%) &        120 (72.29\%) &          48 (96.0\%)  \\
Type &      2 (basement, only)&         5 (1.67\%) &           5 (6.02\%) &            0 (0.0\%) &            0 (0.0\%)  \\
 &        3 (multiple storage)&       60 (20.07\%) &         12 (14.46\%) &         46 (27.71\%) &            2 (4.0\%)  \\
 \\ Storage  &        0 (no)&        17 (5.69\%) &           2 (2.41\%) &          14 (8.43\%) &            1 (2.0\%)  \\
Sufficiency &        1 (yes)&      282 (94.31\%) &         81 (97.59\%) &        152 (91.57\%) &          49 (98.0\%) \\
\\ Shared &        0 (no)&      291 (97.32\%) &         75 (90.36\%) &        166 (100.0\%) &         50 (100.0\%) \\
Storage &        1 (yes)&         8 (2.68\%) &           8 (9.64\%) &            0 (0.0\%) &            0 (0.0\%) \\
\\ Weekly  &      1 ($\leq$ 30) &        10 (3.34\%) &            1 (1.2\%) &            1 (0.6\%) &           8 (16.0\%) \\
 Coverage  &      2 (>30 - 50) &         8 (2.68\%) &            0 (0.0\%) &            2 (1.2\%) &           6 (12.0\%)  \\
 (\% of water &      3 (>50 - 90) &       51 (17.06\%) &           3 (3.61\%) &         31 (18.67\%) &          17 (34.0\%) \\
 demand covered by piped water)&      4 (>90 - 100) &      230 (76.92\%) &         79 (95.18\%) &        132 (79.52\%) &          19 (38.0\%)  \\
\\ Household  &      1 (1 - 4) &      110 (36.79\%) &         19 (22.89\%) &         64 (38.55\%) &          27 (54.0\%) \\
Size &      2 (5 - 7) &      144 (48.16\%) &         36 (43.37\%) &         87 (52.41\%) &          21 (42.0\%) \\
 &      3 ($>$7) &       45 (15.05\%) &         28 (33.73\%) &          15 (9.04\%) &            2 (4.0\%) \\
\\ Income &      1 (>265 - 350) &       81 (27.09\%) &            1 (1.2\%) &         77 (46.39\%) &            3 (6.0\%) \\
(JD/capita) &      2 (>350 - 600) &       74 (24.75\%) &         16 (19.28\%) &         26 (15.66\%) &          32 (64.0\%) \\
  &      3 (>600 - 900) &      106 (35.45\%) &         39 (46.99\%) &         57 (34.34\%) &          10 (20.0\%) \\
 &      4 (>900) &       38 (12.71\%) &         27 (32.53\%) &           6 (3.61\%) &           5 (10.0\%)\\
\\ Water Bill &        1 ($\leq$5)&      196 (65.55\%) &         69 (83.13\%) &        100 (60.24\%) &          27 (54.0\%)  \\
(JD/capita/qtr) &        2 (>5 - 10)&       70 (23.41\%) &          9 (10.84\%) &         47 (28.31\%) &          14 (28.0\%)  \\
 &        3 ($>$10)&       33 (11.04\%) &           5 (6.02\%) &         19 (11.45\%) &           9 (18.0\%)  \\
\\Problems &        0 (no)&      269 (89.97\%) &          82 (98.8\%) &        140 (84.34\%) &          47 (94.0\%)  \\
Paying Bill &        1 (yes)&       30 (10.03\%) &            1 (1.2\%) &         26 (15.66\%) &            3 (6.0\%) \\

\\ Water Social &      1 (0) &      126 (42.14\%) &           7 (8.43\%) &         98 (59.04\%) &          21 (42.0\%) \\
Network &      2 (1-3) &      158 (52.84\%) &         65 (78.31\%) &         66 (39.76\%) &          27 (54.0\%)  \\
 &      3 (4-6) &        15 (5.02\%) &         11 (13.25\%) &            2 (1.2\%) &            2 (4.0\%)  \\
\\Relocation   &      1 (moved within Jordan)&       60 (20.07\%) &         12 (14.46\%) &         18 (10.84\%) &          30 (60.0\%)  \\
 &      2 (moved from abroad)&        27 (9.03\%) &           7 (8.43\%) &          11 (6.63\%) &           9 (18.0\%)  \\
 &      3 (did not move)&      212 (70.90\%) &         64 (77.11\%) &        137 (82.53\%) &          11 (22.0\%)  \\
 \\ Length of Stay &      1 ($\leq$1) &         3 (1.00\%) &           3 (3.61\%) &            0 (0.0\%) &            0 (0.0\%)\\
(years) &      2 (>1 - 5) &        18 (6.02\%) &          9 (10.84\%) &           9 (5.42\%) &            0 (0.0\%) \\
 &      3 ($>$5) &      278 (92.98\%) &         71 (85.54\%) &        157 (94.58\%) &         50 (100.0\%) \\
\\ Housekeeper   &      0 (no)&      266 (88.96\%) &         59 (71.08\%) &        157 (94.58\%) &         50 (100.0\%) \\
 &      1 (yes)&       33 (11.04\%) &         24 (28.92\%) &           9 (5.42\%) &            0 (0.0\%)\\
\\Health &        0 (no)&      273 (91.30\%) &         76 (91.57\%) &        148 (89.16\%) &          49 (98.0\%)  \\
Problems &        1 (yes)&        26 (8.70\%) &           7 (8.43\%) &         18 (10.84\%) &            1 (2.0\%) \\
\\Quality  &        0 (no)&      228 (76.25\%) &         78 (93.98\%) &        112 (67.47\%) &          38 (76.0\%) \\
Problems &        1 (yes)&       71 (23.75\%) &           5 (6.02\%) &         54 (32.53\%) &          12 (24.0\%)  \\
\\District   &      1 &        25 (8.36\%) &           2 (2.41\%) &         23 (13.86\%) &            0 (0.0\%) \\
 &      2 &        25 (8.36\%) &           3 (3.61\%) &         22 (13.25\%) &            0 (0.0\%)  \\
 &        3 &        25 (8.36\%) &         25 (30.12\%) &            0 (0.0\%) &            0 (0.0\%)  \\
 &      4 &        25 (8.36\%) &            1 (1.2\%) &         24 (14.46\%) &            0 (0.0\%) \\
 &      5 &        24 (8.03\%) &         23 (27.71\%) &            1 (0.6\%) &            0 (0.0\%) \\
 &      6 &        25 (8.36\%) &           2 (2.41\%) &         23 (13.86\%) &            0 (0.0\%) \\
 &      7 &        25 (8.36\%) &            1 (1.2\%) &         24 (14.46\%) &            0 (0.0\%)\\
 &      8 &        25 (8.36\%) &            0 (0.0\%) &         25 (15.06\%) &            0 (0.0\%) \\
 &      9 &        25 (8.36\%) &            0 (0.0\%) &            0 (0.0\%) &          25 (50.0\%)  \\
 &      10 &        25 (8.36\%) &            0 (0.0\%) &            0 (0.0\%) &          25 (50.0\%) \\
 &      11 &        25 (8.36\%) &         25 (30.12\%) &            0 (0.0\%) &            0 (0.0\%)  \\
 &        12 &        25 (8.36\%) &            1 (1.2\%) &         24 (14.46\%) &            0 (0.0\%)\\
\end{longtable}

\begin{longtable}{p{0.13\textwidth}p{0.17\textwidth}p{0.13\textwidth}p{0.13\textwidth}p{0.13\textwidth}p{0.13\textwidth}p{0.13\textwidth}}
\toprule
\caption{Outcome Variables: Categories and Cluster-Wise Proportion of Households}\label{tab:2} \\ \\
Outcome Variable &  Category (response) & Overall Count (\%) & Cluster 1 Count (\%) & Cluster 2 Count (\%) & Cluster 3 Count (\%) \\
\midrule
\endfirsthead
\toprule
Outcome Variable &  Category (response) & Overall Count (\%) & Cluster 1 Count (\%) & Cluster 2 Count (\%) & Cluster 3 Count (\%) \\
\midrule
\endhead
\midrule
\endfoot 
\bottomrule
\endlastfoot
Alternate  &      0 (never)&      221 (73.91\%) &         66 (79.52\%) &        150 (90.36\%) &           5 (10.0\%)  \\
Source (AS) &      1 (daily/weekly)&       63 (21.07\%) &         10 (12.05\%) &          10 (6.02\%) &          43 (86.0\%)  \\
Frequency &      2 (monthly)&         7 (2.34\%) &           2 (2.41\%) &           4 (2.41\%) &            1 (2.0\%) \\
&      3 (yearly)&         8 (2.68\%) &           5 (6.02\%) &            2 (1.2\%) &            1 (2.0\%) \\
\\Treat Water  &        0 (no)&       37 (12.37\%) &           5 (6.02\%) &         29 (17.47\%) &            3 (6.0\%)  \\
 &        1 (yes)&      262 (87.63\%) &         78 (93.98\%) &        137 (82.53\%) &          47 (94.0\%) \\
\\Delay Strategy &      0 (no delay)&       40 (13.38\%) &            1 (1.2\%) &         28 (16.87\%) &          11 (22.0\%)  \\
(DS) &      1 (contact utilities)&      147 (49.17\%) &         63 (75.90\%) &         76 (45.78\%) &           8 (16.0\%)  \\
 &      2 (bottled water)&        10 (3.34\%) &            1 (1.2\%) &           4 (2.41\%) &           5 (10.0\%)  \\
 &      3 (borrow)&       33 (11.04\%) &           6 (7.23\%) &         22 (13.25\%) &           5 (10.0\%)  \\
 &      4 (tanker truck)&       59 (19.73\%) &         11 (13.25\%) &         29 (17.47\%) &          19 (38.0\%)  \\
&      5 (additional reserves)&         24 (8.02\%) &            8 (9.63\%) &           15 (9.03\%) &            1 (2.0\%)  \\ 
 &      6 (wait)&         4 (1.34\%) &            1 (1.2\%) &            1 (0.6\%) &            2 (4.0\%)  \\
\\Alternate &      0 (none)&      200 (66.89\%) &         66 (79.52\%) &        127 (76.51\%) &           7 (14.0\%)  \\
Source (AS) &      1 (bottled water)&       70 (23.41\%) &         12 (14.46\%) &           19 (11.44\%) &          39 (78.0\%)  \\
 &      2 (tanker truck)&        42 (14.04\%) &           5 (6.02\%) &         29 (17.46\%) &            8 (16.0\%)  \\
 &      4 (neighbours/well)&         3 (1.00\%) &            0 (0.0\%) &            3 (1.80\%) &            0 (0.0\%)  \\
\end{longtable}

\begin{figure}[H]
    \centering
    \includegraphics[width=18cm]{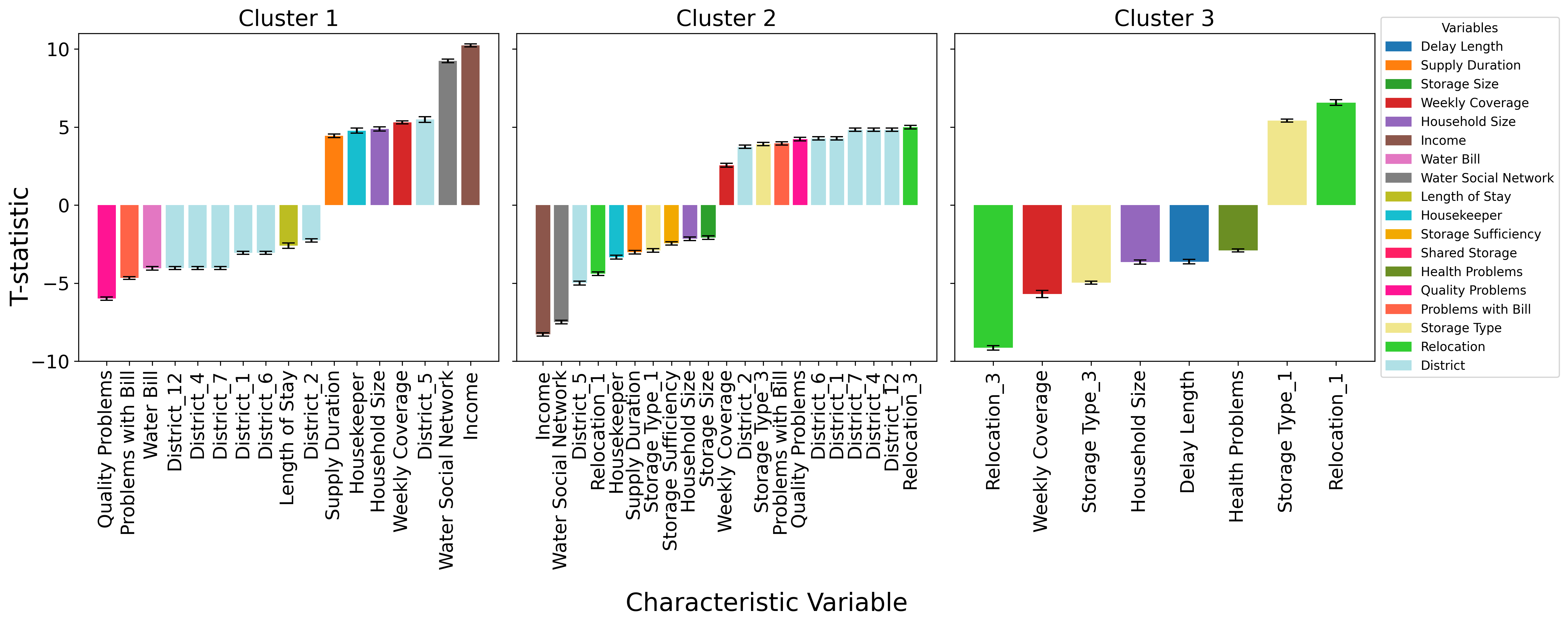}
    \caption{Welch two-sample t-statistic for significant characteristic variables ($\alpha<0.05$) describing clusters 1-3 (left to right). Error bars represent the standard error of the $t$-statistic from the Welch two-sample t-test, indicating the variability in the estimated $t$-values for each variable, which are very low here.}
    \label{fig:3}
\end{figure}

\begin{figure}[H]
    \centering
    \includegraphics[width=18cm]{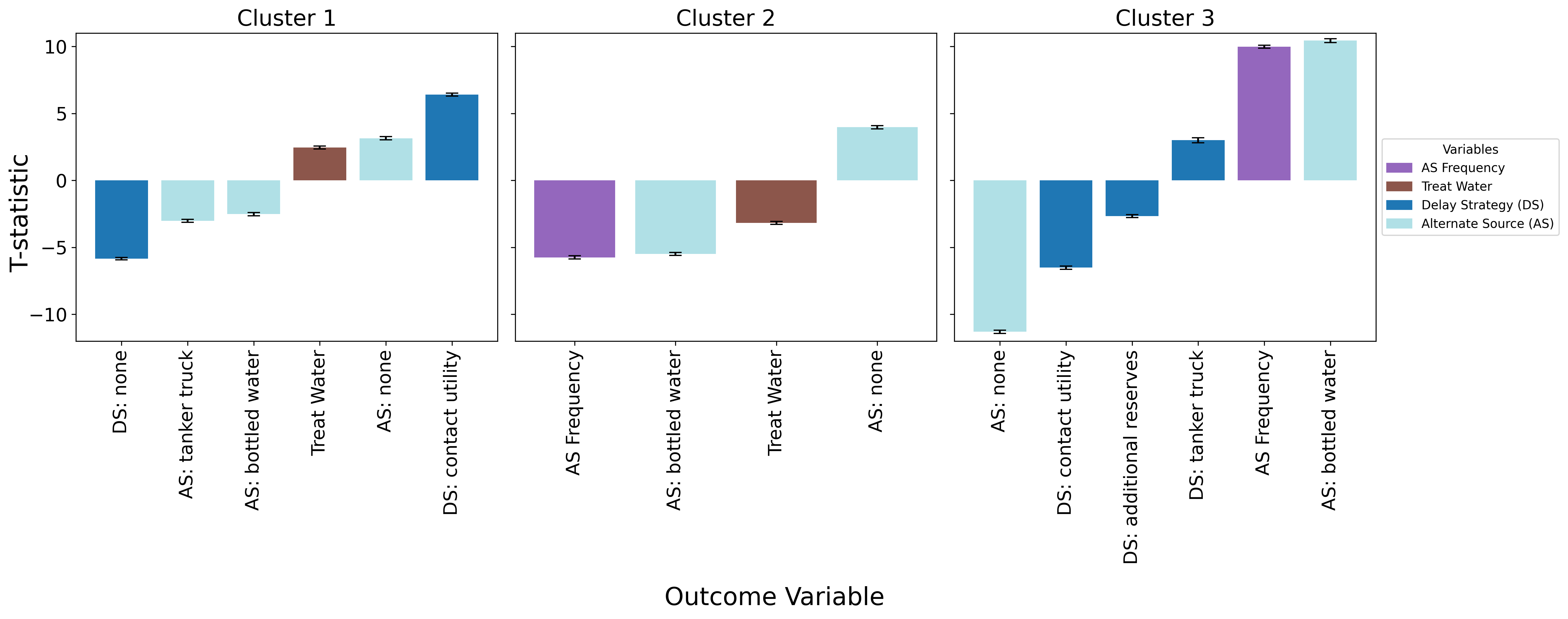}
    \caption{Welch two-sample t-statistic for significant outcome variables ($\alpha<0.05$) describing clusters 1-3 (left to right). Error bars same as in fig \ref{fig:3}.}
    \label{fig:4}
\end{figure}

The three clusters were analyzed and interpreted to uncover which household characteristics are significant ($\alpha<0.05$) in describing each cluster and the subsequent adaptive behaviors using the Welch two-sample t-test, as shown in Figures \ref{fig:3} and \ref{fig:4}, where the $t$-value is plotted for each significant variable across all clusters. Subsequently, Table 3 records variables (categories) with cluster(s) in which household responses have no variance (0\% or 100\%) for that variable (category).

Figure \ref{fig:s5} and \ref{fig:s4} in appendix section \ref{awch} illustrate how each variable (characteristic and outcome, except District) features in each household, displaying the overall heterogeneity of households within each cluster. We used the Welch two-sample t-test to extract distinct features significant for each cluster and used these to describe each cluster.

%\begin{figure}[H]
%    \centering
%    \includegraphics[width=16cm]{draft 3/heatmapout.png}
%    \caption{Heatmaps representing household adaptive behaviors within each cluster. The color bar denotes the category value to which the household belongs for a particular variable. }
%    \label{fig:s5out}
%\end{figure}

\newpage

\begin{longtable}
{p{0.28\textwidth}p{0.20\textwidth}p{0.20\textwidth}p{0.20\textwidth}p{0.18\textwidth}}
\toprule
\caption{Variables (categories) with no variance in household responses in one or more clusters, for which the $t$-test fails\\\centering ($n_{1} = 83$, $n_{2} = 166$, $n_{3} = 50$)}\label{tab:3}\\ \\
Variable (category)&  Cluster 1 (response \%) & Cluster 2 (response \%) & Cluster 3 (response \%) \\
\midrule
\endfirsthead
\toprule
Variable (category)&  Cluster 1 (response \%) & Cluster 2 (response \%) & Cluster 3 (response \%) \\
\midrule
\endhead
\midrule
\endfoot 
\bottomrule
\endlastfoot
Housekeeper: 0 (no)  &      - &       - &          100\% \\
Length of Stay: 3 (>5 yrs)  &      - &      - &         100\% \\
Shared Storage: 1 (yes) &      - &        0\% &         0\% \\
Storage Type: 2 (basement, only) &      - &       0\% &         0\% \\
District: 1 &      - &       - &         0\% \\
District: 2 &      - &       - &         0\% \\
District: 3 &      - &       0\% &         0\% \\
District: 4 &      - &       - &         0\% \\
District: 5 &      - &       - &         0\% \\
District: 6 &      - &       - &         0\% \\
District: 7 &      - &       - &         0\% \\
District: 8 &      0\% &       - &         0\% \\
District: 9 &      0\% &       0\% &         - \\
District: 10 &     0\% &       0\% &         - \\
District: 11 &      - &       0\% &       0\%  \\
District: 12 &      - &       - &         0\% \\
AS: neighbours/well &      0\% &       - &         0\% \\
 \end{longtable}
 
\subsection{Cluster 1} 
Figure \ref{fig:3} shows the $t$-test outcomes for the characteristic variables of households in cluster 1. The figure indicates a strong, significant positive difference between the cluster mean and the remaining data mean for the variables income and water social network, and, to a lesser extent, district 5, weekly coverage, household size, housekeeper, and supply duration. The high $t$-values for the top-two characteristics indicate that this cluster is characterized by a relatively large proportion of households with high income per capita (80\% earn more than 600JD, compared to 48\% of the entire sample), and a relatively large proportion of households equipped with a water social network (91.5\% are equipped with a social network to contact regarding issues with water and about 13\% have relatively large social network size ($\geq$4), compared to only 5\% in the overall data). %Social network, here, comprises friends, relatives, colleagues, or family who help during water supply delays and other issues.  
Their piped water supply covers a relatively large proportion of their weekly water demand (for over 95\% of households, piped water supply covers at least 90\% of water demand compared to 77\% in the entire sample). This cluster also has a higher proportion of large households (34\% have $>$7 members compared to 15\% in the entire sample).  Furthermore, this cluster includes a high proportion of households that have a housekeeper to help collect and store water (29\% compared to 11\% in the entire sample), and with relatively high supply duration (52\% having supply duration 3-7 days/week compared to 35\% in the entire sample).
Figure \ref{fig:3} also shows strong, significant negative $t$-values for water quality problems and problems paying the water bill. This indicates that the majority of households in this cluster does not experience water quality problems (94\% compared to 76\% in the entire sample), nor with paying their water bill (83\% compared to 66\% in the entire sample). Additionally, a high proportion of households has comparatively cheaper water bills (83\% having a water bill of less than 5 JD/capita/qtr, compared to 65\% in the entire sample).

The households in this cluster are spread into districts 3 (30\%), 11 (30\%), and 5 (28\%) as indicated in Table 1, which belong to the center, north, and west parts of Amman, respectively. Table \ref{tab:3} records the variables in this cluster that have no variance in household response either in cluster 1 or in the rest of the data (and subsequently, do not have a meaningful $t$-statistic). 

%These variables are district 3, district 11, Figure \ref{fig:3} does not show $t$-values for districts 3 and 11 as 100\% of households belonging to these districts in the overall data (25 households each) fall in this cluster, making them perfect predictors (see section \ref{ttest}). Other perfect predictors for this cluster are shared storage (8 households) and storage type-2 (basement, only) (5 households).

Adaptive behaviors of households in cluster 1 are shown in Figure \ref{fig:4}. Strong, significant positive $t$-values in this cluster resulted for delay strategy 'DS: contact utilities', alternate source 'AS: none', and 'AS: treat water', while the dominant negative $t$-value resulted for 'DS: no delay'. The latter indicates that, in cluster 1, there was a relatively high proportion of households experiencing water supply delays (76\% of households indicated a delay of at least a day compared to 55\% in the entire sample). In response to delays, the most common strategy was to contact the utility (Miyahuna) (76\% versus 49\% in the entire sample) (fig \ref{fig:4}). The vast majority of households in  cluster 1 treat their water at home (94\%), and they do not access an alternate water source (80\%).

%\begin{figure}[H]
%    \centering
%    \includegraphics[width=15cm]{draft 3/tdf_Cluster_1.png}
%    \caption{Welch two-sample t-statistic for key characteristic variables describing cluster 1. Error bars represent the standard error of the $t$-statistic from the Welch two-sample t-test, indicating the variability in the estimated $t$-values for each variable, which are very low here.}
%    \label{fig:3}
%\end{figure}

%\begin{figure}[H]
%    \centering
%    \includegraphics[width=14cm]{draft 3/tdfo_Cluster_1.png}
%    \caption{Welch two-sample t-statistic for key outcome variables describing   cluster 1. Error bars same as in Figure \ref{fig:3}
%    \label{fig:4}
%\end{figure}

\subsection{Cluster 2}

Figure \ref{fig:3} shows the $t$-test outcomes for the characteristic variables of households in cluster 2. The $t$-values indicate a strong, significant negative difference between the cluster mean and the remaining data mean for the variables income and social network. These values indicate that cluster 2 is characterized by a relatively large proportion of households with low income per capita (46\% earn less than 350JD, compared to 27\% of the entire sample), and a relatively large proportion of households are not equipped with a water social network for help regarding issues with water (60\% of the households report to have no one to contact in case of experiencing water issues compared to 42\% in the overall data).
Figure \ref{fig:3} also shows relatively high positive $t$-values for variables relocation-3 (originally from Amman), districts 1, 2, 4, 6, 7, and 12, quality problems, problems paying bill, and storage type-3 (multiple storage). This indicates that a large proportion of households (82.5\%) in this cluster are originally from Amman, they experience water quality problems (32\% compared to 24\% in the entire sample) and problems with paying their water bill (16\% compared to 10\% in the entire sample), and they have multiple storage (28\% compared to 20\% in the entire sample). 

The households in this cluster are spread into districts 1 (14\%), 2 (13\%), 4 (14\%), 6 (14\%), 7 (14\%), 8 (15\%), and 12 (14\%), which are spread across south-east (1, 2 and 4), center (7, 8 and 12) and north (6) of Amman. Table \ref{tab:3} records the variables in this cluster that have no variance in household response either in cluster 1 or in the rest of the data (and subsequently, do not have a meaningful $t$-statistic). 

 Adaptive behaviors of households in cluster 2 are shown in Figure \ref{fig:4}. The dominant positive $t$-value in this cluster resulted for 'AS: none'. A strong, significant negative $t$-value resulted for 'AS frequency', which indicates a relatively high proportion of households that do not access an alternate source (around 85\%), and for 'AS: bottled water', indicating that these households do not access bottled water as an alternate source (only 4\% of the households use bottled water). Additionally, 'treat water' also shows a moderately negative $t$-value, indicating the cluster contains a larger proportion of households that do not treat water (17\% compared to 12\% in the entire sample).  
 
 %AS: neighbors/well  is a perfect predictor for this cluster since all 3 households in the dataset using neighbors/well as an alternate source belong to this cluster. 

\subsection{Cluster 3}

Figure \ref{fig:3} shows the t-test outcomes for the characteristic variables of households in cluster 3. The t-values indicate a relatively significant positive difference between the cluster mean and the remaining data mean for the variables relocation-1 (moved within Jordan) and storage type-1 (rooftop storage, only). The high positive t-values for these characteristics indicate that this cluster is characterized by a relatively large proportion of households who have moved to Amman from within Jordan (60\% compared to 20\% in the entire sample) and who only have a rooftop storage (96\%). 
Figure \ref{fig:3} also shows relatively high negative t-values for variables relocation-3 (originally from Amman), weekly coverage, storage type-3 (multiple storage), household size, delay length, and health problems. Thus, a relatively large fraction of households in this cluster are originally not from Amman (only 22\% compared to 71\% in the entire sample), few have multiple storage (4\% compared to 20\% in the entire sample), and a large proportion of households in this cluster report low weekly coverage (62\% report weekly coverage below 90\% compared to 23\% in the entire sample), no delays (70\% compared to 44\% in the entire sample), being small households (54\% have household size 1-4 compared to 37\% in the entire sample), and an absence of health problems (98\%). 

Table \ref{tab:3} shows t-values for features without a variance in this cluster and thereby no meaningful t-statistic. Districts 9 and 10 are two of such features as 50\% of households from this cluster  each belong to districts 9 and 10 which are located in the center and west of Amman, respectively. All the households from the entire sample belonging to these districts fall in this cluster. Similarly, other such features for this cluster are housekeeper (none of the households in this cluster have a housekeeper) and length of stay (all households in this cluster have been in Amman for more than 5 years). 

Regarding adaptive behaviors of households shown in Figure \ref{fig:4}, we found that in this cluster dominant positive t-value resulted for 'AS: bottled water' and 'AS frequency'. This indicates that a relatively large proportion of households in this cluster access an alternate source daily or weekly (86\% compared to 21\% in the entire sample) and have bottled water as their primary alternate source (78\% compared to 23\% in the entire sample). Figure \ref{fig:4} also indicates relatively high negative t-values for 'AS: none' and 'DS: contact utility', which indicates that very few households in this cluster have no alternate source (14\%) or contact utilities in case of delays (16\%).

\subsection{Comparison of Household Clusters}

\begin{figure}[H]
    \centering
    \includegraphics[width=15cm]{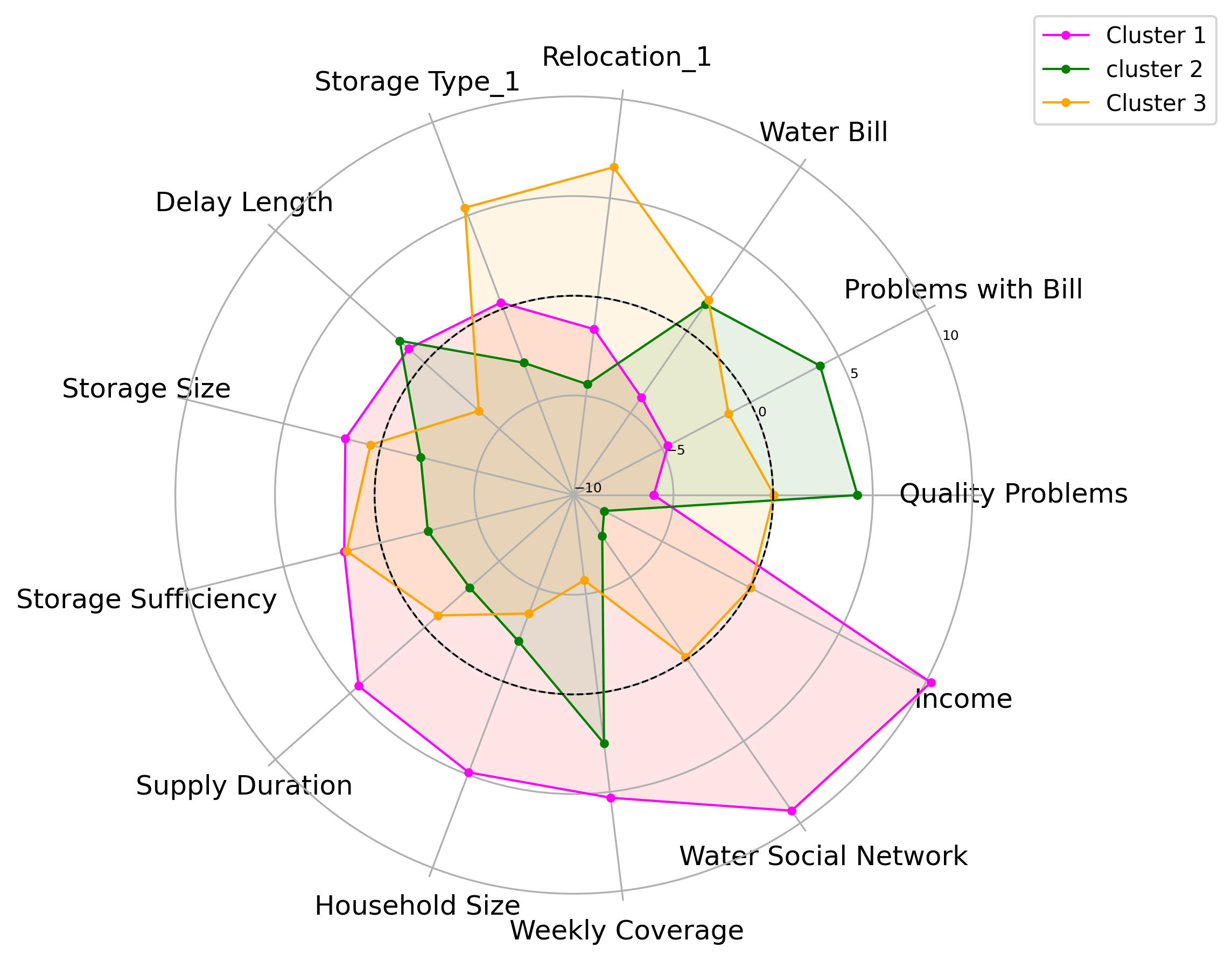}
    \caption{Radar plot of cluster characteristics: $t$-values of significant characteristic variables plotted for each cluster, showing heterogeneity in cluster characteristics}
    \label{fig:5}
\end{figure}

\begin{figure}[H]
    \centering
    \includegraphics[width=14cm]{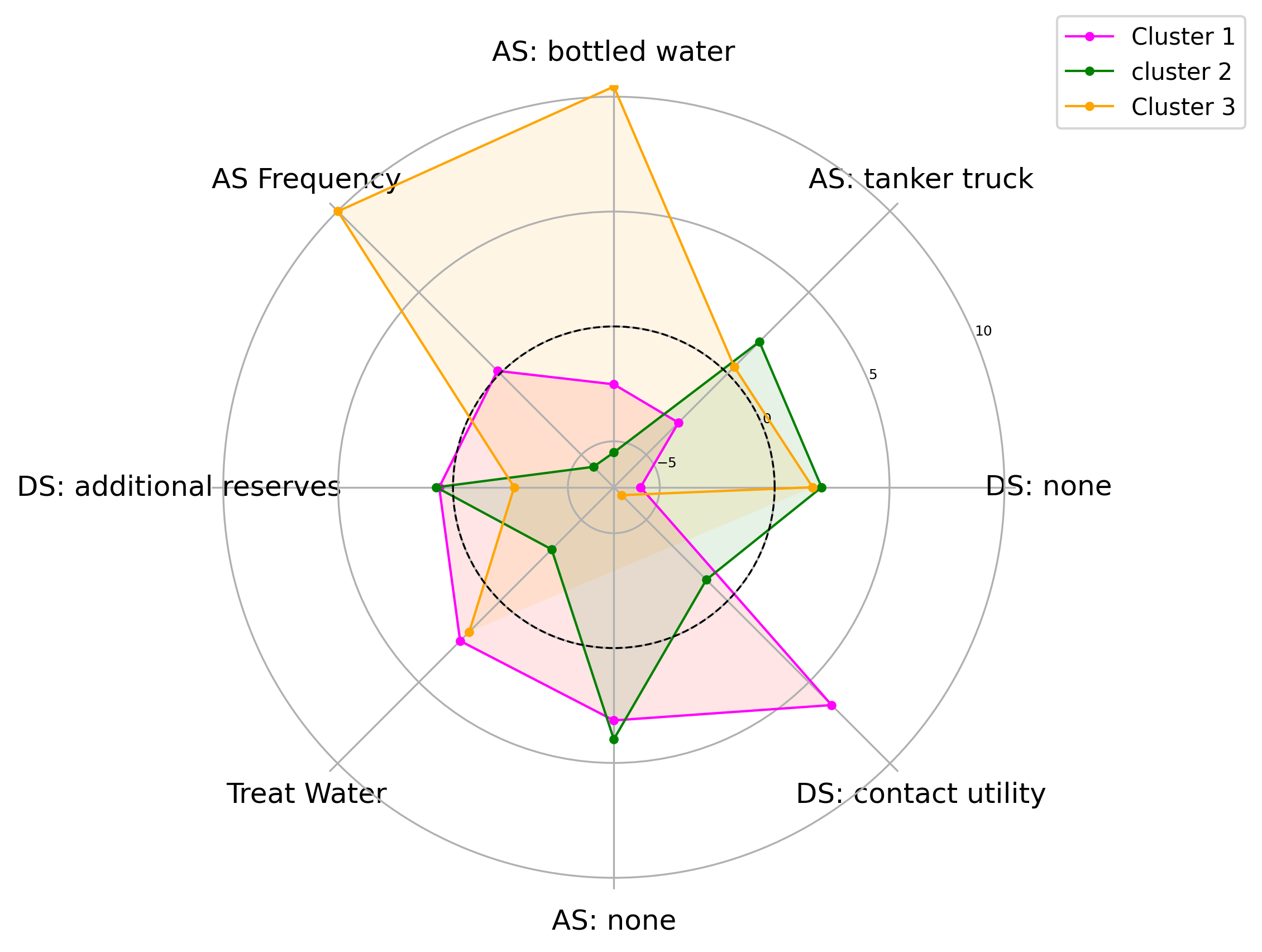}
    \caption{Radar plot of cluster adaptive behaviors: $t$-values of significant outcome variables plotted for each cluster, showing heterogeneity in cluster adaptive behaviors}
    \label{fig:6}
\end{figure}

The results in Figure \ref{fig:5} and \ref{fig:6} compare characteristics and adaptive behaviors of the 3 clusters using radar plots. Each cluster is represented by a polygon with edges representing the t-values for each significant variable, identified by the Welch two-sample t-test (the variables for which a $t$-value exists and have $\alpha<0.05$ after multiple correction). Figure \ref{fig:5} shows the heterogeneity in characteristics across clusters by visualizing the relative difference in the t-values. For example, for income, cluster 1 has a high positive t-value compared to cluster 2, which has a high negative t-value, and cluster 3 has t-value close to 0. This indicates that, relative to the other clusters, cluster 1 is skewed towards higher-income households and cluster 2 towards low-income households. Similarly, Figure \ref{fig:6} shows the heterogeneity in adaptive behaviors across clusters by visualizing the relative difference in the t-values. For example, for AS (alternate source) frequency, cluster 3 has a high positive t-value compared to cluster 2, which has a high negative t-value, and cluster 1 has t-value close to 0. This indicates that, relative to the other clusters, cluster 3 is skewed towards households frequently accessing an alternate source and cluster 2 towards households rarely accessing an alternate source. Cluster-wise polygons indicate the inequality across several characteristics and adaptive behaviors among the household clusters. 

\subsection{Within Cluster Correlations}
While the t-values obtained by the Welch two-sample t-test suggest significant variables differentiating each cluster from rest of the dataset, the clusters are not homogeneous, and are also characterized by within-cluster variability (as shown by the cluster-wise proportions in tables \ref{tab:1} and \ref{tab:2}, and figure \ref{fig:s4} in appendix section \ref{awch}). To demonstrate this, we perform Spearman correlation analysis to analyze which of the significant features within a cluster are correlated. Figures \ref{fig:s6}-\ref{fig:s8} in appendix section \ref{awcc} show pairwise correlations between significant characteristics and outcome variables within each cluster, respectively. For example, cluster 1 is characterized by high income and by long supply duration, however, the negative correlation between these two characteristics within cluster 1 indicates that high income households within this cluster tend to have shorter supply durations, while those with lower incomes tend to have longer supply duration. While all correlations are relatively weak, this is expected, as households are clustered according to distinct features. We also do not observe significant correlations between specific characteristic and outcome variables, which supports our argument that household adaptive behaviors are determined by a set of characteristics that are linked in complex ways, rather than linearly and by individual features.

\section{Discussion}\label{dc}
This paper illustrates a data-driven approach to identify heterogeneous household groups adapting to IWS using household survey data from Amman. We identify three types of household groups distinguished by a set of heterogeneous underlying characteristics, and adapting to IWS in Amman with group-specific behaviors. This advances previous studies beyond single, linear variables, such as income, to explain the unequal nature of water accessibility, consumption, water storage capacity, problems and costs, as well as differences in adaptive strategies such as accessing alternate water sources and contacting the utility to report supply delays.

We uncover the multidimensional nature of social inequality that determines accessibility to water and adoption of water management strategies within IWS for households in Amman. Our analysis shows distinct patterns of adaptation across clusters. Cluster 1 is a set of high income households, equipped with a water social network and high weekly coverage of demand, who don't need to access an alternate source but contact the utility in case of delays. Cluster 2, on the contrary, is a set of low-income households, unequipped with a water social network, experiencing water quality problems and without any significant adaptive strategy. Cluster 3 is a set of households relocated to Amman, equipped only with a rooftop storage and frequently accessing an alternate source, typically bottled water. This hints at the differential adaptive behaviours to IWS of groups of households in Amman, considering multiple underlying characteristics that lead to group-specific strategies, which differentially affect the wellbeing of households within IWS. It also forms a basis for further exploring the effect of factors like social networks and migration with respect to adaptation to IWS.

Previous studies on variations in household water use and management state the need for quantitative methods that are able to reveal intersectional factors and nuanced patterns of adaptation while balancing site-specific outcomes and multi-sited comparison \parencite{shah_variations_2023, galaitsi_intermittent_2016, majuru_how_2016}. The combination of hierarchical clustering analysis (HCA) and Welch two-sample t-test forms an easily interpretable standardized quantitative approach to uncover heterogeneity in IWS adaptation from household survey data, revealing site-specific patterns (for Amman, in this study), while being applicable to uncover heterogeneous household groups across other case studies on IWS. There is also a need for more integrative approaches that consider the multidimensional nature of human-resource interactions in social-ecological systems, generally \cite{pacheco-romero_data-driven_2022}. Hence, this study advances methodologies in capturing the multidimensional nature of human adaptation to service deficits (IWS, in this case), adding to the literature on identifying social-ecological complexity \parencite{khan_beyond_2022, li_intermittentwater_2020, potter_contemporary_2010} \cite{pacheco-romero_data-driven_2022}. 
\\A review on household adaptation to IWS across case studies shows the existence of multiple coping strategies within a city that can be categorized by its purpose, such as enhancing water quality, quantity, or pressure \parencite{majuru_how_2016}. Our study contributes to this literature through a method that identifies group-specific adaptive strategies with different underlying sets of household characteristics. Further distinctions within a cluster can be drawn by examining the subclusters (by lowering the phenon line in the dendrogram, see section \ref{asc}), however this did not result in characterization of distinct subgroups. HCA makes it possible to investigate groups of households at different levels (in the dendrogram) varying in the degree of heterogeneity, however choice of a particular linkage distance depends on the number of data points and available information about the system. For the household survey data in Amman, a linkage distance of 30, identifying three clusters of households resulted in an optimal description of heterogeneity in characteristics and adaptive outcomes to IWS for households in Amman.  

On average, urban residents in Amman received water for 2.5 days a week in 2015, which is reduced to 24 hours a week in 2024 \parencite{krueger_reframing_2025}, clearly indicating the growing problem of insufficient water services exacerbating inequality. Sustainable Development Goals (SDG) target 6.1 lays emphasis on universal and equitable access to safe and affordable drinking water for all by 2030, and to achieve that, accessible and equal water supply is one of the key parameters. Studying heterogeneity of households adapting to IWS is important to achieve an equitable access to water, as SDG is limited to more higher-level (national) indicators. Previous research has shown that local water experts and utility managers in Amman do not adequately account for household efforts and heterogeneity in adaptive strategies to IWS, indicating weak links between them \parencite{krueger_reframing_2025}. In order to mitigate these challenges, it is important to understand the various adaptive strategies and their costs that households adopt to take responsibility for their water in an IWS \parencite{majuru_how_2016}. Adaptive capacity enhances the resilience of households to IWS by enhancing their ability to mobilize water in response to delays or intermittency in supply, which can lessen their vulnerability \parencite{chapagain_studies_2025} but also affect the patterns of consumption. Our analysis can help inform public entities in revising water management policies towards greater equity, who may lack information about the diversified and unequal nature of their citizens' coping behavior \parencite{de_marchis_pressure-discharge_2015, hoekstra_urban_2018}.

\subsection{Limitations and Future Work}
%Experiences of water users have been largely overlooked in several conceptualizations of urban water systems. Urban water systems can be conceptualized as social-ecological-technological systems (SETS) and this paper is a step towards understanding the social element through heterogeneity in water users' adaptation to IWS. Resilience of a household in IWS can be understood as the ability to buffer, persevere, respond and reorganize during periods without water, which is largely determined by its adaptive capacity. While city-scale resilience of urban water supply security has been quantified before, we contribute to this literature by uncovering household resilience against water insecurity through differential adaptation. This study can be useful in developing resilience-oriented and equitable water management that integrates knowledge about different water users into a SETS framework \parencite{krueger_reframing_2025}. 
There are some methodological limitations associated with processing survey data including imputation of missing values. Analysis with bootstrapping is a good alternative, but given the small number of missing values in our dataset (details: appendix), we do not expect this to be an issue. The choice of HCA is suitable for this study but it involves certain assumptions. Methodologically, this study assumes that Euclidean distance (within HCA) is appropriate for all standardized numeric/ordinal variables (including one-hot encoded variables), necessitating that the features used must be suitably chosen to distinguish meaningful differences among households. Also, Ward’s criterion tends to create somewhat spherical (or variance-minimizing) clusters in the standardized feature space. We assume that the variables do not produce extremely elongated or manifold‐like clusters. Alternative methods for clustering in such cases can be used such as self-organised maps (SOMs) \parencite{ioannou_exploring_2021} or density-based clustering methods \parencite{hennig_clustering_2015}. The meaning of data and aim of clustering ultimately determines the selection of a clustering method \parencite{hennig_clustering_2015}. 

This study uncovers clusters of households for a single snapshot in time with correlations exploring within-cluster variability. Future studies that model household adaptation and water use within IWS should explore causal dynamics of adaptation and uncover the differential resilience of households across heterogeneous groups. This would advance previous models on water users within IWS that have only considered households divided in their income levels \parencite{klassert_modeling_2015, rosenberg_modeling_2007, srinivasan_reimagining_2015}. 

Although this study is limited to understanding clustering of household groups adapting to IWS for a single case study, the method used in this analysis can be potentially applied across datasets from other cities facing IWS. We recommend future studies on household surveys in cities with IWS to incorporate clustering methods to uncover a set of socioeconomic determinants and inequality in adaptation across contexts, which can lead to a multi-site comparison of adaptation to IWS \parencite{majuru_how_2016, galaitsi_intermittent_2016, simukonda_intermittent_2018}. Future data collection could benefit from shared protocols that contribute to the compatibility for clustering analysis. 

%human behavior, causal-models and dynamics, subcluster analysis-within cluster heterogeneity, collecting data

\section{Conclusion}\label{cc}
This is a first study that explores a standardized clustering method to uncover the heterogeneity in socio-economic determinants of household adaptation to IWS and to identify group-specific adaptive strategies. Previous studies have stated household income to be the main determinant of coping strategies adopted against IWS \parencite{majuru_how_2016}. However, our analysis demonstrates the complex nature of socio-economic factors underlying IWS adaptation, giving insights into the different significant factors that can characterize household groups within a city with IWS. Specifically for Amman, we arrive at three household clusters with differential adaptive capacities. 

%To summarize, the main features of household clusters in Amman: 
%\begin{itemize}
%    \item Cluster 1: High income households equipped with social network and high weekly coverage. They do not access alternate sources however their main delay strategy is to %contact utility.
%    \item Cluster 2: Low income households not equipped with social network who experience problems with water quality and paying bills on time. They have no alternate sources or delay strategies.
%    \item Cluster 3: Households migrated to Amman with only a rooftop storage and low weekly coverage. They access alternate sources more frequently, with bottled water being one of the most common source.
%\end{itemize}
This analysis also links a set of socio-economic characteristics like income, water social network, supply duration, relocation and water quality problems to group-specific adaptive behaviors like contacting the water utility or accessing an alternate water source, contributing to a better understanding of heterogeneity within urban IWS. This study introduces a standardized method: Step 1: Hierarchical Clustering Analysis, Step 2: Welch two-sample t-test, Step 3: Identification of adaptive behaviors. This method has advantages in revealing context-specific or site-specific heterogeneity within IWS adaptation while potentially being applicable across other IWS contexts, enabling more meaningful cross-site comparisons. This requires standardized survey protocols to allow for comparative analyses. The locally grounded insights can support urban water managers in understanding differential adaptation among water users within IWS, leading to interventions that account for the varied realities and constraints faced by urban households \parencite{de_marchis_pressure-discharge_2015}.

\printbibliography
%\addbibresource{references1.bib}

\appendix
\section{Data Pre-processing}\label{adp}
\subsection{Missing Values}\label{amm}
Missing values and ambiguous responses (for example, 'I don't know') in the dataset were handled with the following steps: (a) For some of the ordinal variables (number of such responses) (supply duration (68), delay length (22), and water bill (27)), such a response was assumed to be in the most favorable category related to IWS adaptation, like, long supply duration and no delays (b) For water bill (27) and storage size (5), such responses were imputed using the mean to preserve central tendency before transforming them into ordinal categories, (c) For water social network (39), such responses were assumed from households who do not resort to social network for water issues, and were placed in category with water social network size: 0, (d) (appendix) For nominal variable length of stay, there was one ambiguous response: 'she did not wish to answer', which was assumed to be in category 1 ($\leq$ 1 yr), and the corresponding missing response under the variable relocation, was assumed to be in category 2 (moved from abroad). Another missing response in relocation variable was assumed to be in category 3 (did not move), looking at the corresponding length of stay to be more than 5 years, (e) For nominal variables alternate source and delay strategy, such a response was assumed to be associated with no alternate source or no delays and was placed in the corresponding category (f) For some binary variables (health problems (6), quality problems (20) and problems with bill (3)), ambiguous responses were either inferred into a yes/no category based on the response (for example, 'colour and smell' in was placed in the yes category for quality problems), whereas missing responses were imputed in category 0 (no problems), towards which the data was skewed in each of these columns. 

\subsection{Correlation Analysis}\label{acorr}
\begin{figure}[H]
    \centering
    \includegraphics[width=15cm]{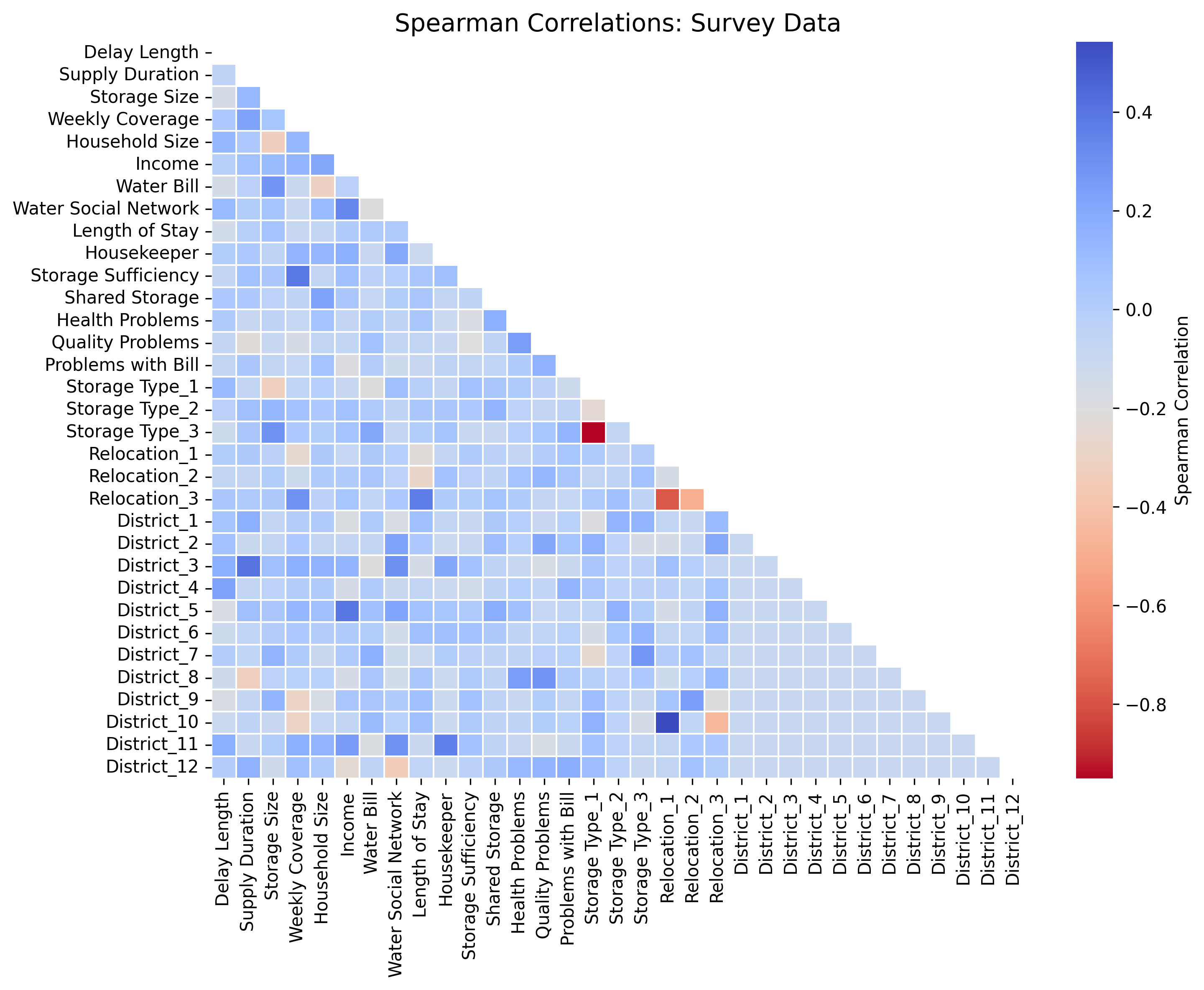}
    \caption{Pairwise correlation between chosen variables in the overall dataset: The colour represents spearman's correlation coefficient}
    \label{fig:s1}
\end{figure}

We performed Spearman's Correlation Analysis to compute pairwise correlation between chosen variables in the overall household survey data and Figure \ref{fig:s1} plots the correlation coefficient ($\rho$) between pairs of variables, by colour. We observed most of the variables not correlated or weakly correlated with each other. There were three pairs of variables, relatively strongly correlated ($|\rho|>0.5$):
\begin{itemize}
    \item Storage Type-1 and Storage Type-3 ($\rho=-0.95$), indicating that households with only a rooftop storage (strage type-1), do not have multiple storage (storage type-3)
    \item Relocation-1 (moved within Jordan) and Relocation-3 (did not move) ($\rho=-0.78$).
    \item Relocation-1 (moved within Jordan) and District-10 ($\rho=0.54$), indicating a correlation between households that have moved to Amman within Jordan and households living in District-10.
\end{itemize}
The relatively strong correlations in the overall dataset were only found for one-hot encoded variables, which were categories within the same variable (relocation or storage type) stating obvious correlations, except for the correlation between relocation-1 and District-10, which wasn't considered significant enough to remove either variable categories from the analysis as both variables convey different information. 

\subsection{PCA}\label{apca}
\begin{figure}[H]
    \centering
    \includegraphics[width=15cm]{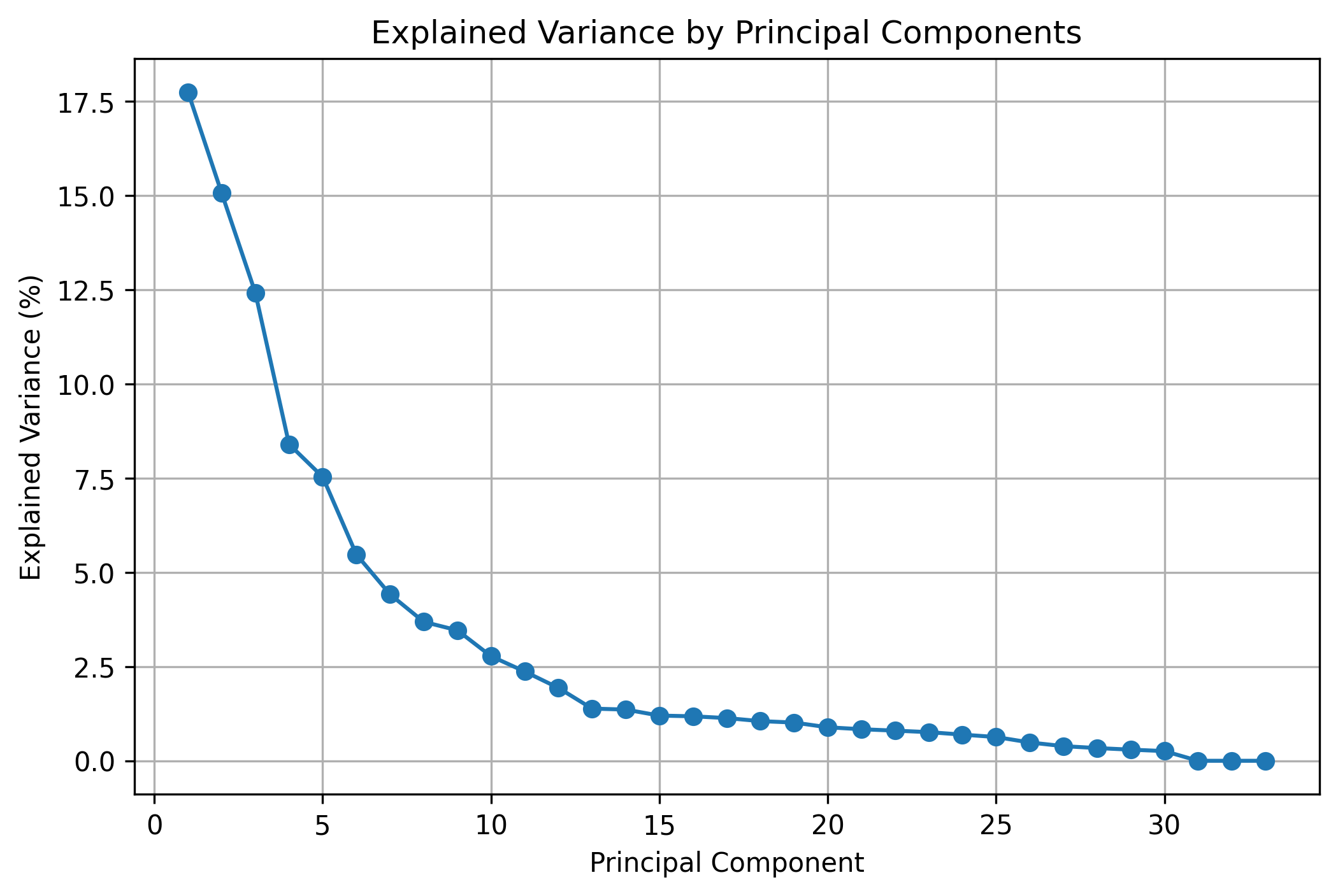}
    \caption{Plot showing explained variance by each principal component after Principal Component Analysis}
    \label{fig:s2}
\end{figure}

\begin{figure}[H]
    \centering
    \includegraphics[width=15cm]{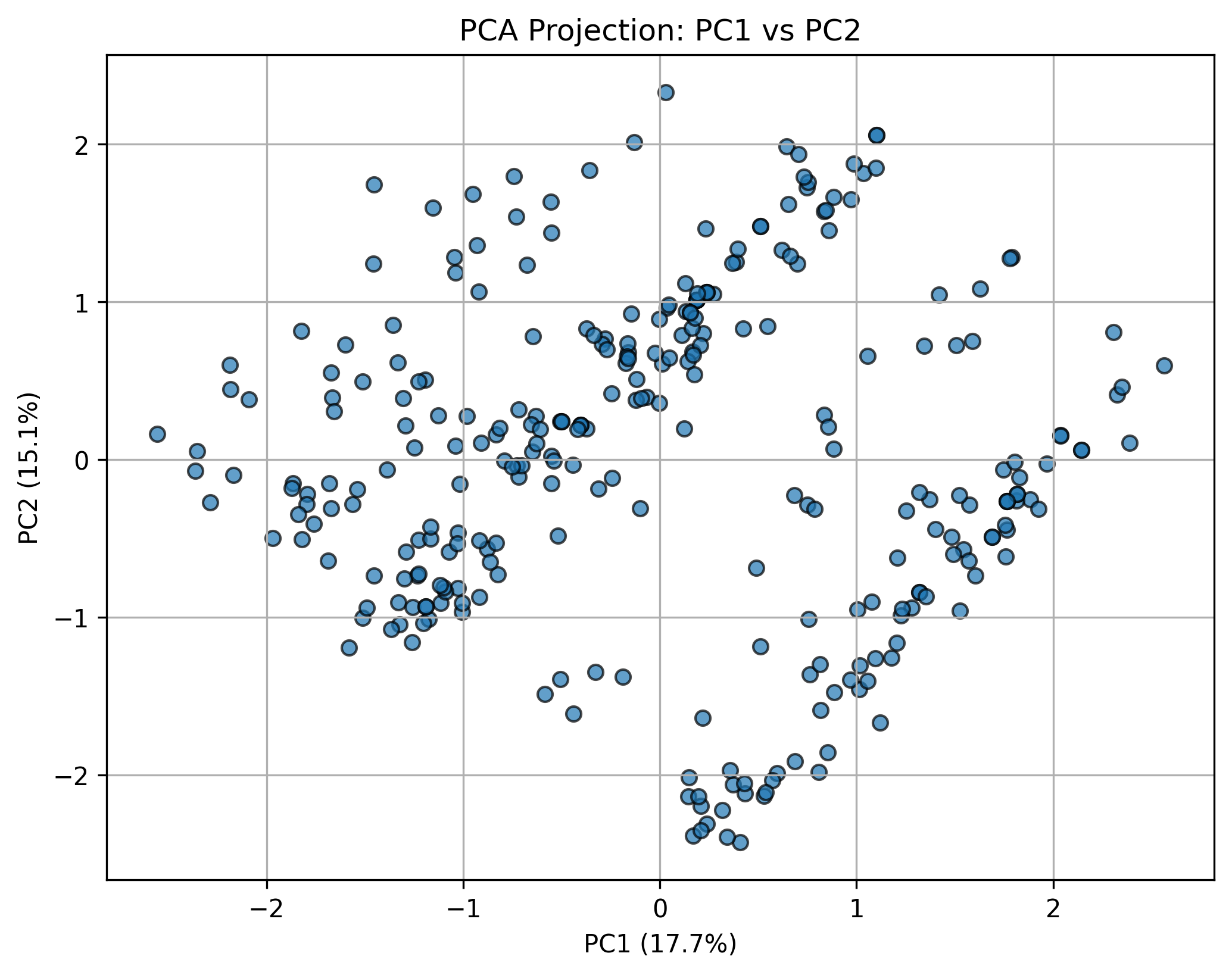}
    \caption{Data points from the overall dataset plotted on a two-dimensional plot with PC-1 and PC-2 as the axes}
    \label{fig:s3}
\end{figure}

Principal Component Analysis (PCA) was performed on the overall dataset and Figure \ref{fig:s2} shows the variance in the dataset explained by each principal component obtained for the data. We observe the first two principal components (PC-1 and PC-2) only explaining about 17.5\% and 15\% variance in the dataset each, with the variance explained by further principal components decreasing gradually. We also observe in figure \ref{fig:s3}, each data point plotted on a two-dimensional plot with PC-1 and PC-2 as the axes and it does not help us characterize significant grouping within the dataset, based on these principal components. Hence, PCA does not prove to be useful in understanding the heterogeneity within this dataset.

\section{Within Cluster Heterogeneity}\label{awch}
\begin{figure}[H]
    \centering
    \includegraphics[width=10cm]{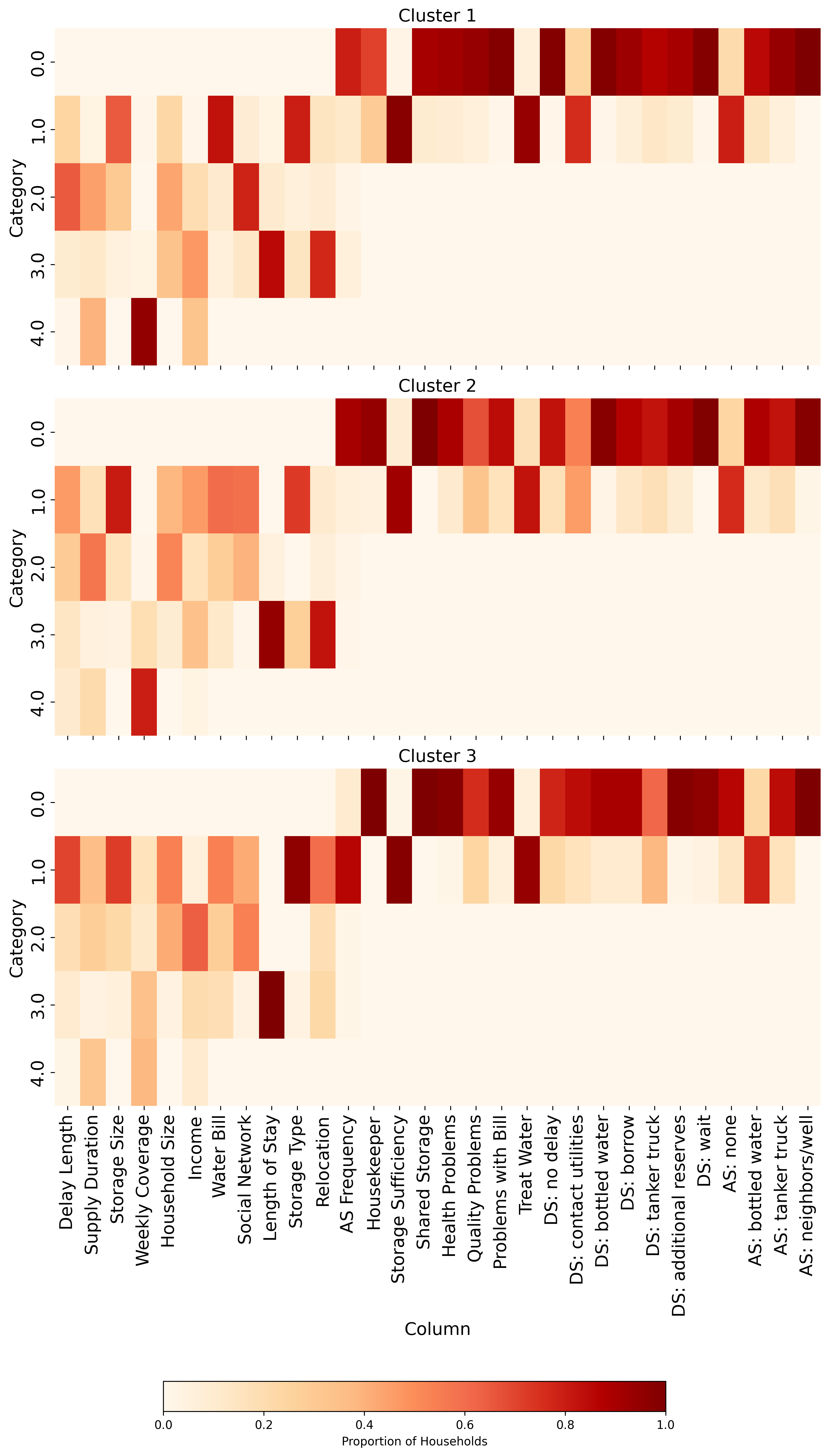}
    \caption{Heatmap representing the distribution of households across variable categories (characteristic and outcome) within clusters 1, 2 and 3 (top to bottom). Variable 'District' with 12 categories is not plotted in this figure.}
    \label{fig:s4}
\end{figure}

\begin{figure}[H]
    \centering
    \includegraphics[width=16cm]{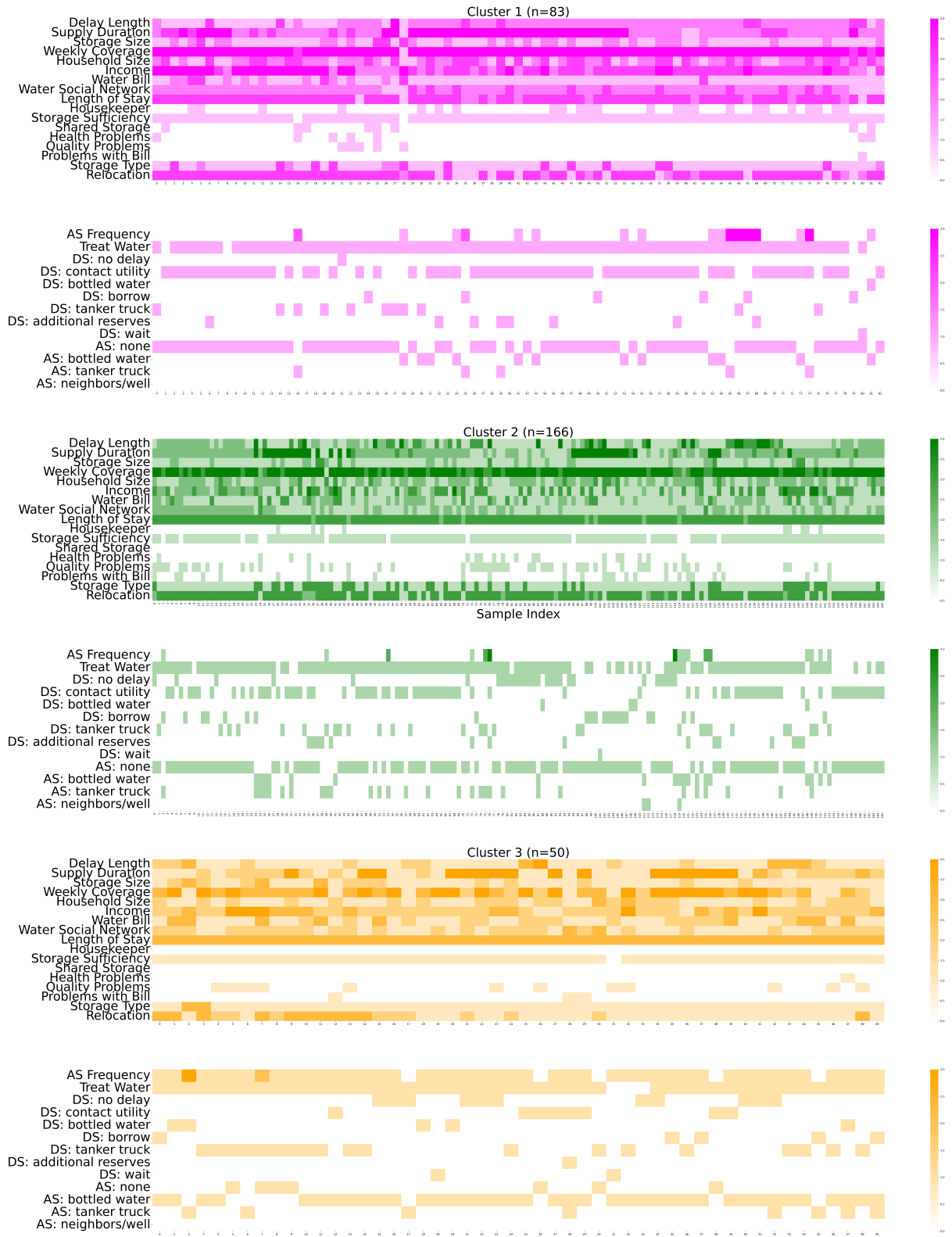}
    \caption{Heatmaps representing household responses to each variable within each cluster (denoted by the color). The intensity of the colorbar denotes the category value to which the household belongs for a particular variable. See Table 1 for an explanation of category values.}
    \label{fig:s5}
\end{figure}

Each cluster is distinguished by a unique set of characteristics and adaptive behaviors that are shown by a relatively large proportion of households within the cluster, compared to rest of the households, as reflected by the $t$-values. However, the household proportions within clusters (figure \ref{fig:s4} and \ref{fig:s5}) alone suggest sufficient within-cluster heterogeneity.

Figure \ref{fig:s4}, show the distribution of households across variable categories, within each cluster. These highlight key patterns within each cluster and illustrate how different variables and adaptive outcomes are distributed across the households within a cluster.
Figure \ref{fig:s5} in appendix section \ref{awch} illustrates how each variable (characteristic and outcome, except District) features in each household, displaying the overall heterogeneity of households within each cluster. 

\section{Within Cluster Correlations}\label{awcc}
For this analysis, we merge the one-hot encoded variables into the original variables (for example, relocation and storage type) and order the variables in the figure according to their significance in that cluster (top to bottom ordered by most positive to most negative t-values for the variable within the cluster). 

\begin{figure}[H]
    \centering
    \includegraphics[width=14cm]{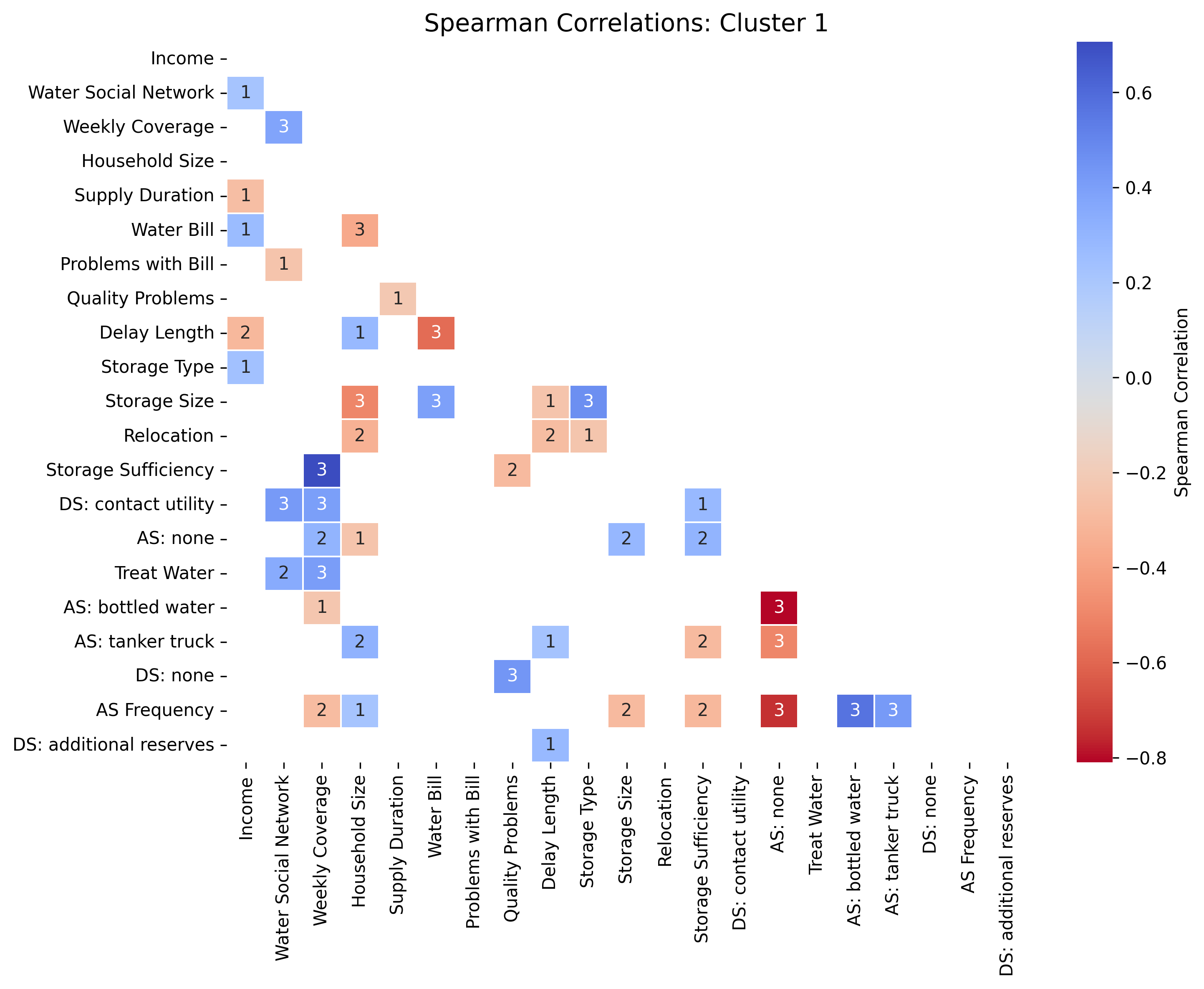}
    \caption{Pairwise correlation between significant variables (characteristic and outcome) within cluster 1: The colour represents spearman's correlation coefficient and the annotated number represents the significance of correlation based on the p-value (1-$*$, 2-$**$, 3-$***$)}
    \label{fig:s6}
\end{figure}

\begin{figure}[H]
    \centering
    \includegraphics[width=15cm]{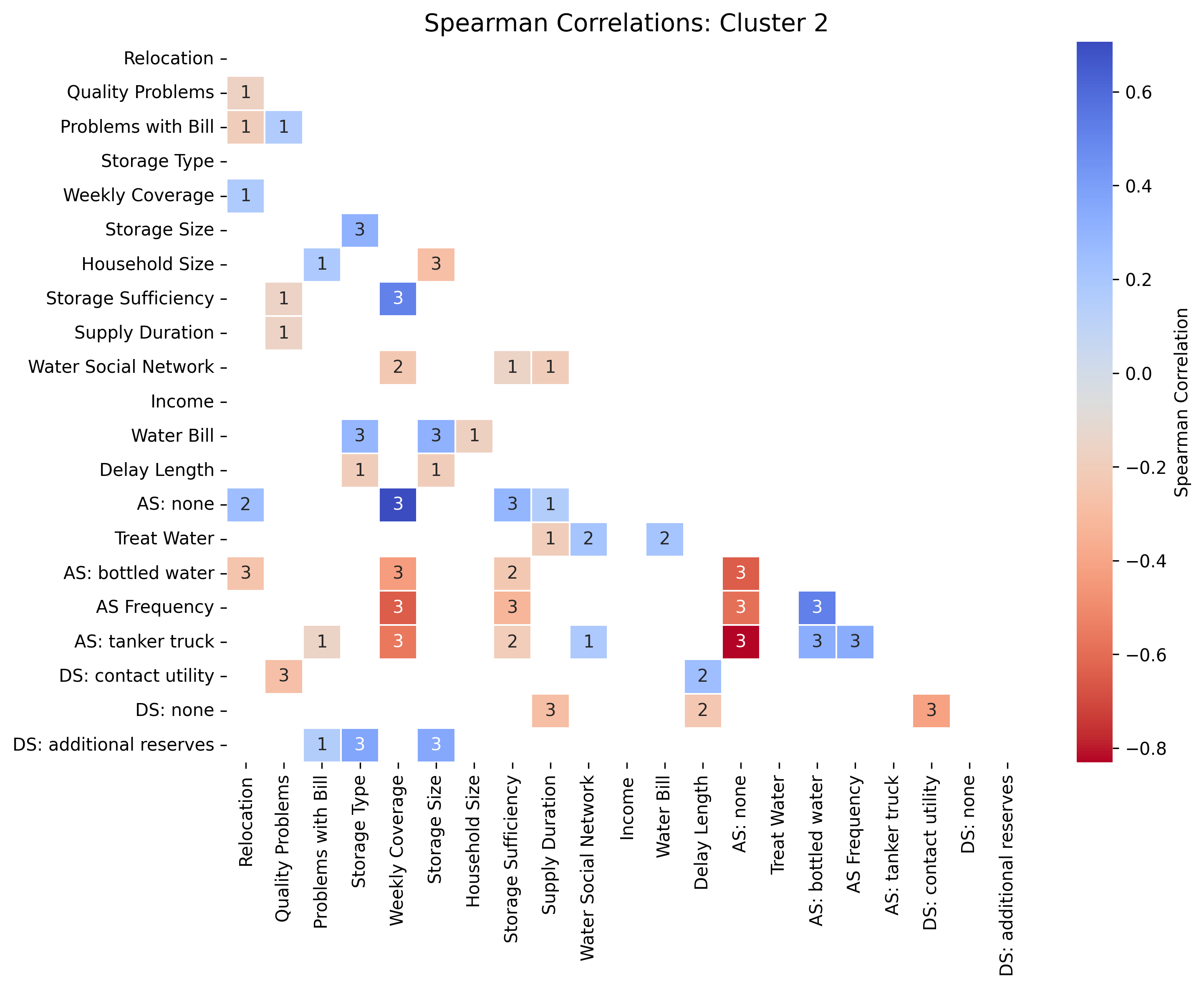}
    \caption{Pairwise correlation between significant variables (characteristic and outcome) within cluster 2: The colour represents spearman's correlation coefficient and the annotated number represents the significance of correlation based on the p-value (1-$*$, 2-$**$, 3-$***$)}
    \label{fig:s7}
\end{figure}

\begin{figure}[H]
    \centering
    \includegraphics[width=15cm]{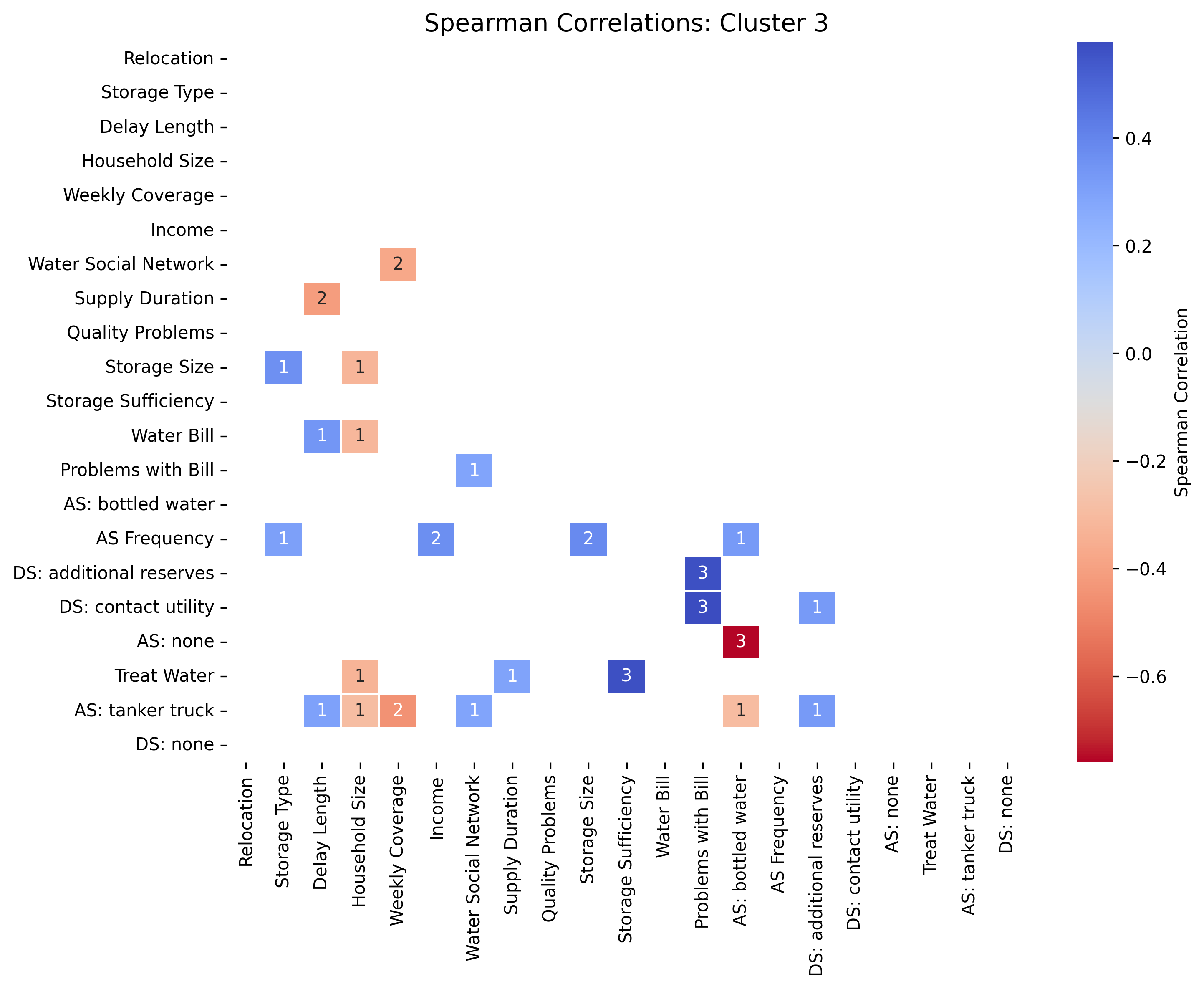}
    \caption{Pairwise correlation between significant variables (characteristic and outcome) within cluster 3: The colour represents spearman's correlation coefficient and the annotated number represents the significance of correlation based on the p-value (1-$*$, 2-$**$, 3-$***$)}
    \label{fig:s8}
\end{figure}

\section{Subcluster}\label{asc}
HCA creates dendrogram that records the merging history of clusters, which allows us to choose any linkage distance to cut the dendrogram (and place the phenon line), and examine the clusters at that merging stage. We choose linkage distance as 30 in our study to observe interpretable clusters. We further lower the linkage distance to 25 to cut the dendrogram, to observe 9 subclusters and comment on their interpretability.

\begin{figure}[H]
    \centering
    \includegraphics[width=15cm]{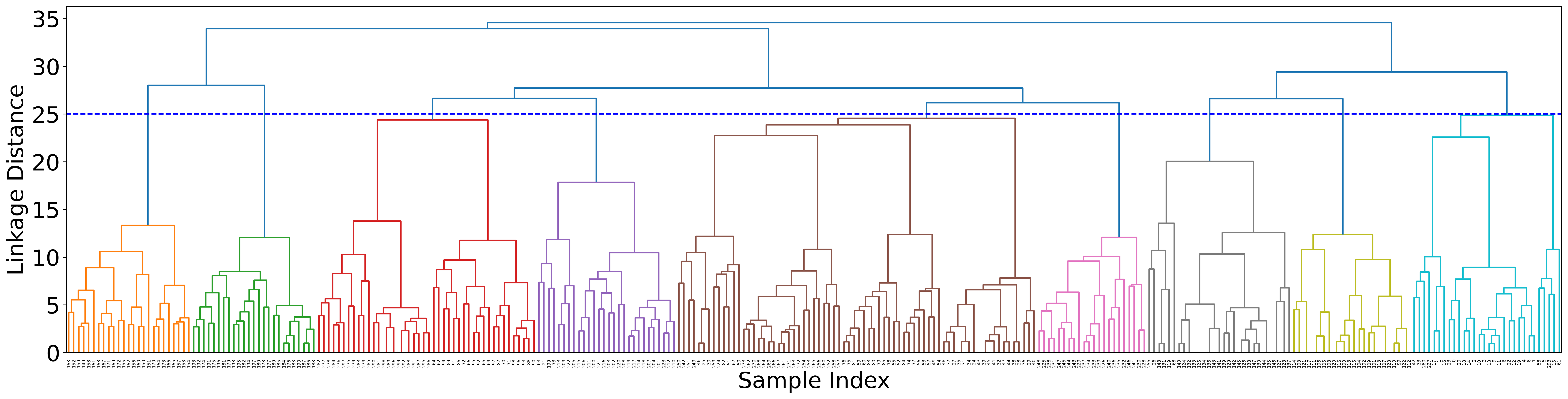}
    \caption{HCA Dendrogram with cluster merging history of household survey data from Amman. Cut at a linkage distance of 25, it results in nine  subclusters (from right to left: subclusters 1 to 9}
    \label{fig:s9}
\end{figure}

The above dendrogram, shows 9 subclusters of households in Amman, which are further groupings within the 3 clusters studied in this paper. We can observe subclusters 1, 2, 3 as groupings within cluster 1; subclusters 4, 5, 6, 7 as groupings within cluster 2; and subclusters 8 and 9 and groupings within cluster 3. We further perform the Welch two-sample t-test on characteristic and outcome variables for each subcluster and look at the significant factors for each of them. 

\begin{figure}[H]
    \centering
    \includegraphics[width=15cm]{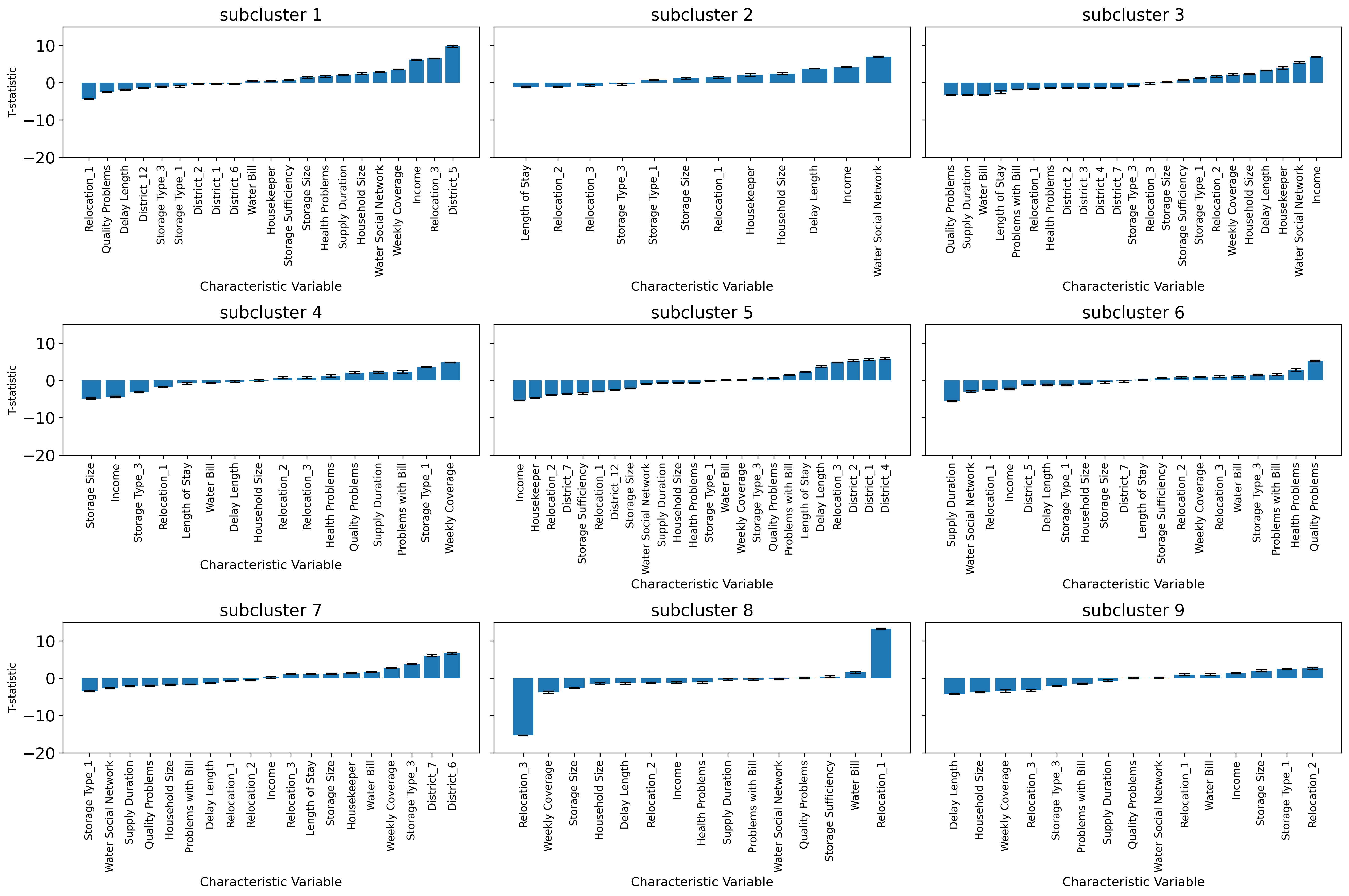}
    \caption{Welch two-sample t-statistic for significant characteristic variables describing subclusters 1-9, with error bars as standard deviation}
    \label{fig:s10}
\end{figure}

\begin{figure}[H]
    \centering
    \includegraphics[width=15cm]{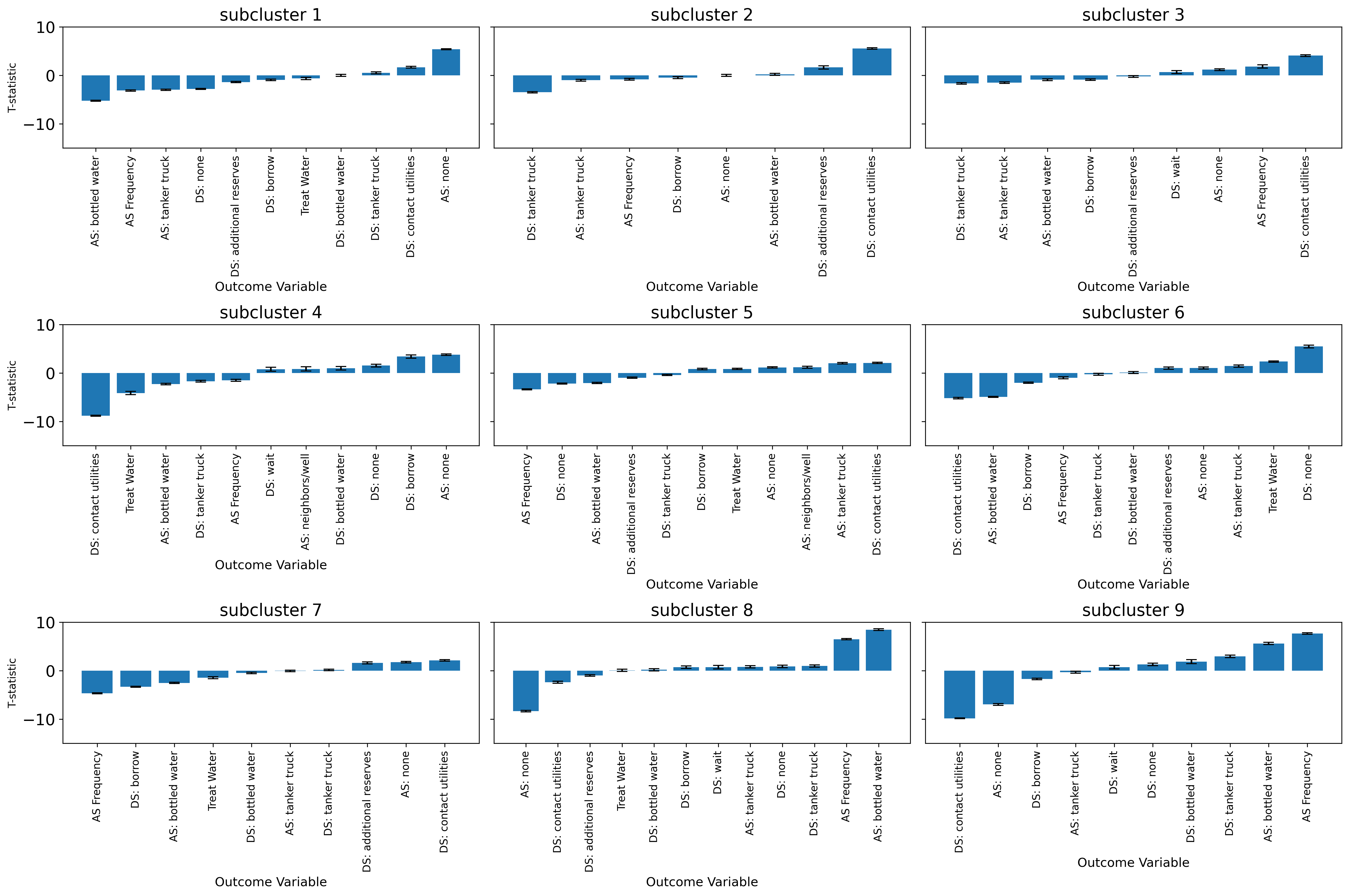}
    \caption{Welch two-sample t-statistic for significant outcome variables describing describing subclusters 1-9, with error bars as standard deviation}
    \label{fig:s11}
\end{figure}

On studying the significant factors characterizing each subcluster within a cluster, we see some subclusters having a specific property of the cluster, while most of the subclusters retain the characteristics of their clusters. For instance, subcluster 1 has significant characteristic as original residents of Amman, subcluster 2 has water social network while subcluster 3 has income. Within Cluster 1, which had income and water social network as most significant characteristics, we see subgroups with one of these as the most significant characteristic. This can be helpful in understanding the demographic more and study the within-cluster heterogeneity. However, it only leads to understanding the spread of characteristics within a cluster and does not lead to characterization of groups. 
\end{document}